\documentclass[12pt,aps,%
               pra,%
               onecolumn,%
               showpacs,%
               superscriptaddress,%
               nofootinbib,%
               floatfix]{revtex4-1}
\usepackage{amsmath,graphicx,color,dsfont}
\usepackage{siunitx}
\usepackage{commath} 
\usepackage{amssymb}
\usepackage{float}
\usepackage{setspace}
\usepackage[utf8]{inputenc}
\usepackage{float}
\usepackage[colorlinks,linkcolor=blue,anchorcolor=blue,linktocpage=true,urlcolor=blue,citecolor=blue]{hyperref}
\begin{document}

\title{%
  Quantitative Study on Circular Dichroism Induction from Achiral Nanostructure-Chiral Matter Near Field Interactions%
}

\author{ Abuduwaili. Abudukelimu}
\email{Abduwelee@snnu.edu.cn}
\author{ Ziyan Zhang}
\affiliation{ {School of Physics and Information Technology, Shaanxi Normal University, Xi'an, 710119, China}}
\author{ Tursunay. Yibibulla}
\affiliation{School of Physics and Electronics, Central South University, Changsha, 410083, P. R. China}
\author{ Muhammad. Ikram}
\affiliation{ {School of Physics and Information Technology, Shaanxi Normal University, Xi'an, 710119, China}}

\begin{abstract}
\setcounter{equation}{0}
\setcounter{figure}{0}
\setcounter{section}{0}
\setcounter{subsection}{0}
\begin{center}
\textbf{Abstract}
\end{center}
 \begin{spacing}{1.2}
The circular dichroism (ICD) induced from interaction of chiral molecules with achiral plasmonic metal nanostructures has improved sensitivity of molecular chirality  detection. Although experimental results have presented several orders magnitude of ICD  enhancement, but the cause of ICD induction has not arise from previous studies. In this paper, a theory presented for explain mechanism of ICD induction from near field interaction of chiral molecule with achiral nanostructure. By introducing dipole approaching of chiral molecule and a near field approximation of molecule-nanostructure
interaction, we derived an analytical formula for ICD. This theory explains mechanism of ICD is that the dipole radiation of chiral molecules breaks the enantiomorphic relationship of the electric field strengths in the nanostructure. Furthermore, the derived formula exhibits ICD proportional to molecules chirality parameter and number density, and it exponentially increases with the inverse cube of distance between molecule and nanostructure. Most importantly, the formula shows that ICD is proportional to strengths of electric field in nanostructure and magnetic field in molecular region.  To verify the analytical results, we studied the ICD properties
of graphene nanohole metasurface (GNM) and chiral molecule composed system. Numerical results were found consistent with analytical results. Our investigation results are
helpful for understanding mechanism of ICD induction,  pushing the limit of ICD enhancement and realizing supersensitive detection of molecular
chirality.

\textbf{Keywords}: Induced circular dichroism, Surface plasmon, Chiral molecules, dipole approaching, near field approximation.
\end{spacing}
\end{abstract}
\maketitle

\section{Introduction}
\label{sec:introduction}

Circular dichroism (CD), absorption difference of right-handed circularly polarized (RCP) and left-handed circularly polarized (LCP) light, is widely used in determining handiness and absolute
conformations of chiral molecules \cite{Fasman, Tullius, YZhao, Besteiro,Maoz}, and enantioselective separation of chiral molecules \cite{Aitzol2017,Aitzol2019}. However, CD of chiral molecules (MCD) is weak and only observable in the ultraviolet (UV) region, the long exposure to UV radiation can severely affect chiral molecular samples. Therefore, it is desired to develop new techniques to increase the limit of molecular chirality sensitivity or transfer CD signal to long wavelength, so as to achieve the ultimate goal of single-molecule chirality detection. Thus the
surface plasmon enhancing technique has been used for enhancing the MCD \cite{Hendry}. In the interaction between chiral molecules and nanostructure, on one hand, the
optical chiral field enhanced by surface plasmon increases the MCD  in the UV region \cite{Schaferling2016}, on the other hand, the molecular chirality
transfers to nanostructure and generates the induced circular dichroism (ICD) signal \cite{YZhao,Hendry}. The ICD provides a way to detect chiral
molecules in the visual or microwave region, which is easily detectable by CD spectroscopy.

In order to further enhance the ICD signal, effects of different factors such as molecules location \cite{BHan,dawei}, molecules density \cite{rongyao,jianjia},
molecular orientation \cite{Govorov2010,TLeviBelenkova}, separation distance between molecule and nanostructure \cite{Maoz}, and shape of nanostructure \cite{YZhao,ehk,ZZhu2012} on
ICD have been investigated. The studies show that ICD is proportional to the near field strength in molecular region \cite{MLNesterov}, thus ICD is larger when chiral molecule is placed on ends than placed on sides for nanorod-chiral molecule coupled system \cite{SHou2014,SHou2015,BHan} due to the stronger near field at end of nanorod. Furthermore, ICD is sensitive to slightly shape change of nanostructure \cite{ehk}, becouse shape changes lead to near field changes.
Despite the properties of ICD is gradually being clear, but mechanism of ICD induction is need to investigate.  And the above studies suggests the cause of ICD induction is strongly associated with near field characteristics such as electric field strength, magnetic field strength and near field distribution.
Although previous studies have obtained the origin of CD enhancement in chiral molecule-nanostructure interaction based on chiral Poynting's theorem \cite{SLee}, but cause of ICD induction has not explained. The molecule-plasmon Coulomb interaction model correctly reproduces ICD through incident light electric field strength and size parameters of nanostructure \cite{Govorov2010,AOGovorov2012}, but it can not also explain the cause of ICD induction due to ICD is strongly associated with characteristics of  near field.
Therefore, quantitative
studies on near field interactions between chiral molecule and nanostructure, and find mechanism of chirality transfer from chiral molecule to achiral structure is desirable.
Specifically, quantitatively studying the effects of electric field strength, magnetic field strength, molecular chirality parameters, molecular-nanostructure distance and molecular density on ICD to discover the mechanism of circular dichroism induction is helpful to improve the sensitivity of molecular chirality detection.

In this paper, a theory based on chiral molecular dipole approaching and molecule-nanostructure near field interaction approximation is presented to analyze the ICD
induction in chiral molecule-achiral nanostructure interaction, and we use the theory to explain the mechanism of ICD induction in achiral nanostructure.
To prove the theory, a chiral molecule-graphene nanohole metasurface (GNM) composed system is proposed for simulation. The simulated results are consistent with the analytical
results. These results would be helpful for realizing supersensitive detection of molecular chirality.
\section{Theory}
Absorption of nanostructure is formally calculated by volume integral $\frac{\omega}{2}\textrm{Im}(\varepsilon)\oint_V |\textrm{\textrm{E}}|^2 dV$\, \cite{SLee,AOGovorov2012}.
In presence of chiral molecule, electric field strength in nanostructure comes from the contributions of the electric field generated by surface plasmons without chiral molecules and the electric field
radiated by chiral molecular dipole. Thus, absorption of the nanostructure under circularly polarized light (CPL) is given by \cite{BHan,zwx}
\begin{equation}
\label{absorption}
  {A}_{\pm}^{ns}=\frac{\omega}{2}\mathrm{Im}(\varepsilon_{ns})\int_{V_{ns}}({\mathrm{\textbf{E}}}_{\pm}^{0}+
  {\mathrm{\textbf{E}}}_{\pm}^{d})^{\ast}\cdot({\mathrm{\textbf{E}}}_{\pm}^{0}+ {\mathrm{\textbf{E}}}_{\pm}^{d})dV,
\end{equation}
where, ${\mathrm{\textbf{E}}}_{\pm}^{0}$ (+ represents RCP light, and - represents LCP light) is the electric field strength inside the nanostructure without the chiral molecule,
${\mathrm{\textbf{E}}}_{\pm}^{d}$ is the dipole radiated electric field strength in nanostructure, ${A}_{\pm}^{ns}$ is the absorption of nanostructure, $\omega$ is the angular frequency of light,
$\varepsilon_{ns}$, $V_{ns}$ are  the permittivity and volume of nanostructure.

The electric fields ${\mathrm{\textbf{E}}}_{\pm}^{d}$ can be written in terms of dyadic Green's function ${\mathrm{\textbf{E}}}_{\pm}^{d}=i\omega
n_{mol}{\overline{\overline{G}}_{\pm}s\left(r\right)}\textbf{P}_{\pm}^{mol}$, where, $n_{mol}$  is the molecules number density, $\textbf{P}_{\pm}^{mol}$ is dipole strength of chiral
molecule,  and $\overline{\overline{G}}_{\pm}\left(r\right)$ is the dyadic Green's function which is the electric field corresponding to molecular dipole at arbitrary location. The Green
function $\overline{\overline{G}}\left(r\right)$ of chiral medium is given by \cite{IVLindell}
\begin{eqnarray}
\label{Green}
\overline{\overline{G}}\left(r\right)&=&\overline{\overline{G}}_{+}\left(r\right)+\overline{\overline{G}}_{-}\left(r\right),\\
\overline{\overline{G}}_{\pm}\left(r\right)&=&-\frac{i\eta}{2}\left[k_\pm\mathrm{G}_{\pm}\left(r\right)\mathds{I}\pm \nabla\mathrm{G}_{\pm}\left(r\right)\times
\mathds{I}
+\nabla\nabla\left(\frac{\mathrm{G}_{\pm}\left(r\right)}{k_\pm}\right)\right]\nonumber,
\end{eqnarray}
where, $\eta=\sqrt{\frac{\mu}{\varepsilon}}$ is the electrical impedance of chiral medium, $\varepsilon$, $\mu$ are permittivity and magnetic permeability of chiral medium,
$\mathds{I}$ is unit dyadic, $r$ is the distance between molecule and location of the observed electric field, and the function $\mathrm{G}_{\pm}\left(r\right)$ is equal to $\mathrm{G}_{\pm}\left(r\right)=\frac{1}{4\pi r} e^{-ik_\pm r}$. Here, $k_\pm=(k\pm\frac{\kappa}{n_c})$ are wave vectors of RCP and LCP, $k$ is wave vector of linear polarized light, $n_c$ is refractive index of chiral molecule.
The chiral molecule and nanostructure are interact through near filed, thus implementing the near-field interaction approximation $k_\pm r\ll1$ to calculate the gradients of function
$\mathrm{G}_{\pm}\left(r\right)$, we derive
\begin{eqnarray}\label{gradiant}
  \nabla \mathrm{G}_{\pm}\left(r\right)&=&\left(i k_\pm+\frac{1}{r}\right)\mathrm{G}_{\pm}\left(r\right){\hat{u}_r}\approx \frac{1}{r}\mathrm{G}_{\pm}\left(r\right){\hat{u}_r},\nonumber\\
  \nabla\nabla \mathrm{G}_{\pm}\left(r\right) &=&\left [\left(i k_\pm+\frac{1}{r}\right)^2+\frac{1}{r^2}\right]\mathrm{G}_{\pm}\left(r\right){\hat{u}_r}{\hat{u}_r} -\frac{1}{r^2}\mathrm{G}_{\pm}\left(r\right)\left(\mathds{I}-{\hat{u}_r}{\hat{u}_r}\right)\\
  &\approx&\frac{3}{r^2}\mathrm{G}_{\pm}\left(r\right){\hat{u}_r}{\hat{u}_r} -\frac{1}{r^2}\mathrm{G}_{\pm}\left(r\right)\mathds{I},\nonumber
\end{eqnarray}
where, ${\hat{u}_r}{\hat{u}_r}$ is a diadic and  ${\hat{u}_r}$ is a unit vector along distance $\textbf{r}$. Replacing $\nabla \mathrm{G}_{\pm}\left(r\right)$ and $\nabla\nabla \mathrm{G}_{\pm}\left(r\right)$
by their respective relations from Eq. (\ref{gradiant}) and implementing near field approximation $k_\pm r\ll1$, Eq. (\ref{Green}) can be written as
\begin{equation}\label{Green1}
\overline{\overline{G}}_{\pm}\left(r\right)= \frac{-i\eta}{2r^2k_\pm} \left[3\mathrm{G}_{\pm}\left(r\right){\hat{u}_r}{\hat{u}_r}-\mathrm{G}_{\pm}\left(r\right)\mathds{I}\right].
\end{equation}

Dipole moment of chiral molecule is given by \cite{LDBarron,YCTang}
\begin{equation}\label{dipole}
  \textbf{P}_{\pm}^{mol}=\alpha{\mathrm{\textbf{E}}}_{\pm}^c-i\frac{\kappa}{c}{\mathrm{\textbf{B}}}_{\pm}^c,
\end{equation}
where, $\mathrm{\textbf{E}}_\pm^c$ and $\mathrm{\textbf{B}}_\pm^c$  are respectively electric field strength and magnetic flux density in molecular region, $\alpha$
and $\kappa$ are respectively the electric polarizability and chirality parameter of chiral molecules, and $c$ is the speed of light.

Finally, we derive the dipole radiated electric field ${\mathrm{\textbf{E}}}_{\pm}^{d}$
\begin{eqnarray}\label{ed}
  \mathrm{\textbf{E}}_{\pm}^{d}&= &i\omega n_mol \alpha \overline{\overline{G}}{\mathrm{\textbf{E}}}_{\pm}^c +\frac{\omega n_mol
  \kappa}{c}\overline{\overline{G}}{\mathrm{\textbf{B}}}_{\pm}^c\nonumber\\
  &=&\frac{\omega\eta n_{mol}\alpha}{2r^2k}G\left(3\hat{u}_r \hat{u}_r-\mathds{I}\right)\cdot
  {\mathrm{\textbf{E}}}_{\pm}^c -\frac{i\omega\eta n_{mol}}{2r^2k}\frac{\kappa}{c}G\left(3\hat{u}_r \hat{u}_r-\mathds{I}\right)\cdot {\mathrm{\textbf{B}}}_{\pm}^c.
\end{eqnarray}
where, $G_{+}\left(r\right)+G_{-}\left(r\right)=G$.

 In order to derive the ICD, we first calculate the field intensity difference in nanostrcture
\begin{eqnarray}
\label{ediffe}
  (\textbf{E}_+^0+\textbf{E}_+^d)\cdot(\textbf{E}_+^0+\textbf{E}_+^d)^*-(\textbf{E}_-^0+\textbf{E}_-^d)\cdot(\textbf{E}_-^0-\textbf{E}_-^d)^* =2\textrm{Re}\left[(\textbf{E}_+^{0*}\cdot\textbf{E}_+^d)-(\textbf{E}_-^{0*}\cdot\textbf{E}_-^d)\right].
\end{eqnarray}

Replacing $\textbf{E}_\pm^d$ to calculate the $(\textbf{E}_\pm^{0*}\cdot\textbf{E}_\pm^d)$, we derive
\begin{eqnarray}
\label{emed}
 &&(\textbf{E}_\pm^{0*}\cdot\textbf{E}_\pm^d)=\textbf{E}_\pm^{0*}\cdot\left[\frac{\omega\eta n_{mol}\alpha}{2r^2k}G\left(3\hat{u}_r
  \hat{u}_r-\mathds{I}\right){\mathrm{\textbf{E}}}_{\pm}^c
  +\frac{-i\omega\eta n_{mol}}{2r^2k}\frac{\kappa}{c}G\left(3\hat{u}_r
  \hat{u}_r-\mathds{I}\right){\mathrm{\textbf{B}}}_{\pm}^c \right] \\
  &&=\frac{3\omega\eta n_{mol}\alpha}{2r^2k}\frac{1}{r^2}G\sum_\nu\sum_\gamma x_\nu x_\gamma\textrm{E}_\pm^{0\nu*}\textrm{E}_\pm^{c\gamma}
 -\frac{\omega\eta
  n_{mol}\alpha}{2r^2k}G\sum_\nu\textrm{E}_\pm^{0\nu*}\textrm{E}_\pm^{c\nu}\nonumber\\
 &\;&-\frac{3i\omega\eta n_{mol}\kappa}{2r^2kc}\frac{1}{r^2}G\sum_\nu\sum_\gamma x_\nu x_\gamma\textrm{E}_\pm^{0\nu *} \textrm{B}_\pm^{c\gamma}
    +\frac{i\omega\eta
 n_{mol\kappa}}{2r^2kc}G\sum_\nu\textrm{E}_\pm^{0\nu*}\textrm{B}_\pm^{c\nu},\nonumber
\end{eqnarray}
Where, $\nu, \gamma=x, y, z$.

Suppose the incident light is propagate along $z$ direction, and the achiral structure is symmetric along $y$ axis. The near field in any $xy$ plain has properties
$\textrm{E}_+^x(x,y)=\textrm{E}_-^x(x,-y)$, $\textrm{E}_+^y(x,y)=\textrm{-E}_-^y(x,-y)$, $\textrm{E}_+^z(x,y)=\textrm{E}_-^z(x,-y)$, $\textrm{B}_+^x(x,y)=\textrm{-B}_-^x(x,-y)$,
$\textrm{B}_+^y(x,y)=\textrm{B}_-^y(x,-y)$, $\textrm{B}_+^z(x,y)=\textrm{-B}_-^z(x,-y)$. Implementing these properties to calculate Eq. (\ref{emed}), the field intensity difference in nanostrcture becomes
 \begin{eqnarray}
 \label{SS15}
\textbf{E}_+^0\cdot\textbf{E}_+^{d\ast}-\textbf{E}_-^0\cdot\textbf{E}_-^{d\ast}&=&\frac{-3i\omega\eta n_{mol}}{r^2k}\frac{\kappa}{c}G\frac{1}{r^2}\sum_\nu x_\nu^2\textrm{E}_+^{0\nu*}\textrm{B}_+^{c\nu}+ \frac{i\omega\eta n_{mol}}{r^2k}\frac{\kappa}{c}G\sum_\nu\textrm{E}_+^{0\nu*}\textrm{B}_+^{c\nu}.
\end{eqnarray}

For periodic structure in $x$ and $y$ directions, Eq. (\ref{SS15}) can approximate as
\begin{eqnarray}
  &&(\textbf{E}_+^0+\textbf{E}_+^d)\cdot (\textbf{E}_+^0+\textbf{E}_+^d)^* -(\textbf{E}_-^0+\textbf{E}_-^d)\cdot (\textbf{E}_-^0-\textbf{E}_-^d)^* \nonumber\\
  &=&\frac{\eta
  n_{mol}}{\pi r^3}\textrm{ Re} \left[-i\kappa e^{-ikr}\left(\textrm{E}_+^{0x*} \textrm{B}_+^{cx}+\textrm{E}_+^{0y*} \textrm{B}_+^{cy}\right)\right].
\end{eqnarray}

Finally, the ICD generated by a molecular dipole strength $\textbf{P}$ is can be written as
\begin{eqnarray}\label{cd}
  \textsc{ICD}=\frac{\eta n_{mol}}{2\pi}\mathrm{Im}(\varepsilon_{m})
  \int_{V_{ns}}\frac{1}{r^3}\textrm{ Re}\left[-i\kappa e^{-ikr}\left(\textrm{E}_+^{0x*}
  \textrm{B}_+^{cx}+\textrm{E}_+^{0y*} \textrm{B}_+^{cy}\right)\right]dV.
\end{eqnarray}

According to Eq. (\ref{cd}), there are linear relationship between the ICD and number density of chiral molecules and chirality parameter of chiral molecules. Moreover, the ICD exponentially increases with the inverse cube of distance between chiral molecules and nanostructure. Most importantly, ICD proportional to electric field
strength in nanostructure and magnetic flux density in chiral molecular region. These results not only show relationships between ICD and near field, this theory also can explain the cause of ICD induction (Section \ref{rad}).
\begin{figure}[b]
\centering
\includegraphics[height=6cm]{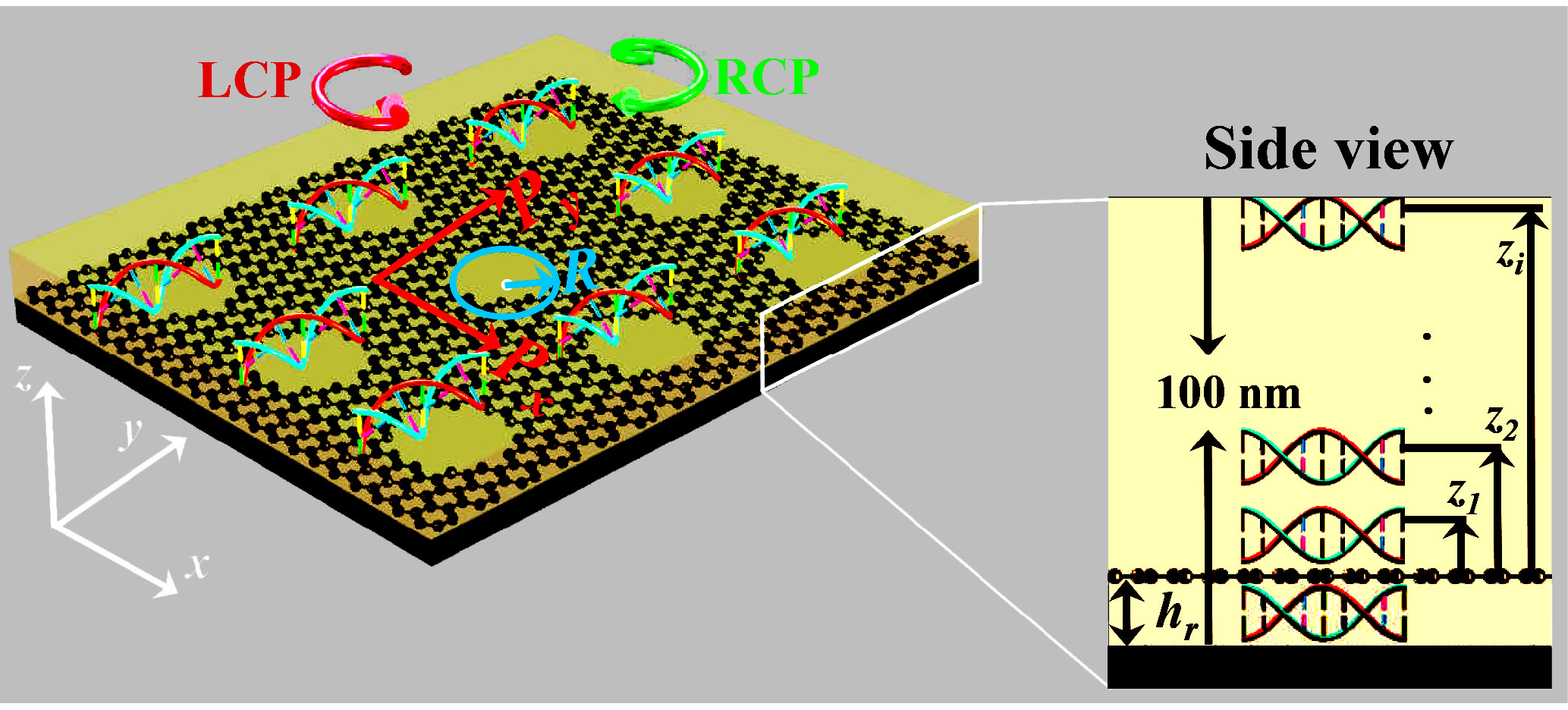}
\caption{Schematic drawing of chiral molecule-GNM composed system.}
\label{fig11}
\end{figure}
\section{Structure and Computational Method }
In order to prove Eq. (\ref{cd}), we propose a chiral molecule and graphene nanohole metasurface (GNM) composed system. Generally, the SPR wavelength of graphene is located in
microwave region, thus the condition $k_\pm r\ll1$  is completely satisfied. Figure \ref{fig11}(a) depicts the proposed chiral molecule-GNM composed system.
The GNM was supposed to be immersed in the bottom of $100\,nm$ thick chiral molecule solution. The periods were fixed as $P_{x}=P_{y}=600\,nm$, and radius of nanohole was chosen to be $150\,nm$. Generally, the chiral molecules have absorption and MCD peak located in UV region \cite{Fasman}, for example the $\alpha$-helix has an absorption and MCD peak located at $192\,nm$.
Therefore, the chiral molecules are modeled with complex refractive index and chirality parameter that has absorption and MCD peak at $192\,nm$.

Finite element method (FEM) was used to perform the simulations. To simulate for chiral medium, Maxwell equations were modulated by implementing a chiral constitutive relation given by \cite{AHSihvola}
\begin{eqnarray}
  \textbf{D}&=&\varepsilon_0 \varepsilon_{c}{{\textbf{E}}}-\frac{i\kappa}{c} {\mathrm{\textbf{H}}},\nonumber \\
  \textbf{B}&=&\mu_0 \mu_{c}{{\textbf{H}}}+\frac{i\kappa}{c} {\mathrm{\textbf{E}}},
\end{eqnarray}
where, $\varepsilon_{c}$, $\mu_{c}$ are the relative permittivity and  magnetic permeability of the chiral molecule, $\kappa$ is the molecular chirality parameter, $\varepsilon_0$
and $ \mu_0$ are the vacuum permittivity and magnetic permeability, respectively. The permittivity $\varepsilon_{c}$ and molecular chirality parameter $\kappa$  are given by
\cite{LDBarron}
\begin{eqnarray}
  \varepsilon_{c} &=& \varepsilon_{sol}+n_{mol}\omega_0 d \frac{\omega_0^2-\omega^2+i\omega\Gamma}{\left(\omega_0^2-\omega^2\right)^2+\omega^2\Gamma^2},\nonumber\\
   \kappa &=& n_{mol}\omega R_{rot} \frac{\omega_0^2-\omega^2+i\omega\Gamma}{\left(\omega_0^2-\omega^2\right)^2+\omega^2\Gamma^2},
\end{eqnarray}
where, $\varepsilon_{sol}$ is the relative permittivity of solvent, $d$ represents the electric dipole strength, $R_{rot}$  is the rotatory strength, $\omega_0$ is molecular absorption
band frequency. The parameters were set as $\varepsilon_{sol}=1.77$, $d = 1.6\times10^{-9}m^3s^{-1}$, $R_{rot} = 1.6\times10^{-11}m^3s^{-1}$, $n_{mol} = 5\times10^{22}m^{-3}$,
$\Gamma = 3.77\times10^{14}s^{-1}$, $\omega_{0} = 9.8\times10^{15}\mathrm{Hz}$. The chirality parameter of a chiral molecule is represented by a matrix. The initial molecular orientation is defined as parallel to $z$-axis, and the unitary
transformation $\kappa_{f}=R(\varphi)\kappa_{i}R^{-1}(\varphi)$  is used for changing of molecular orientation, where,  $\kappa_{i}$ is the initial orientation and $\kappa_{f}$  is
the final orientation, $R(\varphi)$ is unit rotation matrix for angle $\varphi$ which is different for rotations about $x$, $y$ and $z$-axis \cite{AAbudukelimu}.

The conductivity of graphene was computed within the local random phase approximation, which is a function of the frequency of incident light \cite{EHHwang,GWHanson}
\begin{eqnarray}
\sigma(\omega)&=&\frac{2ik_{B}T}{ \pi{\hbar}^2\left(\omega+i \tau^{-1}\right)}\ln\left[2 \cosh\left(\frac{E_f}{2k_BT}\right)\right]\\
&+&\frac{e^2}{4\pi\hbar^2}\left\{0.5+\frac{1}{\pi}\arctan\left(\frac{-\hbar\omega +2E_f}{2k_BT}\right)
-\frac{i}{2\pi}\ln\left[\frac{\left(\hbar\omega+2E_f\right)^2}{\left(\hbar\omega-2E_f\right)^2+(2k_BT)^2}\right]\right\},\nonumber
\end{eqnarray}
where, $T$ is temperature, $\tau$ is carrier relaxation time, $k_B$ is Boltzmann constant, $v_F$ is Fermi velocity, $k_BT$ and $E_f$ represent temperature energy and Fermi energy,
respectively. The parameters were set as $T=300\,K$, $E_f=0.6\,eV$, $v_{F}=c/300$ ($c$ is speed of light), $\tau=\frac{{\rho}_{DC}E_{f}}{e{v}_{F}^2}$ and ${\rho}_{DC}=10^4\,cm/Vs$.

\section{Results and Discussion}\label{rad}
\textbf{Mechanism of circular dichroism induction.} According to Eq. (\ref{absorption}) and Eq.  (\ref{ediffe}), ICD comes from different electric field strengths in nanostructure under RCP and LCP illumination. Here, we use our theory to explain the mechanism of generating the different electric field strengths.

The racemic mixture of chiral molecule does not possess chiral properties and it cannot induce ICD. The radiated electric field by the racemic mixture molecular dipole is
\begin{equation}
\label{edr}
  \mathrm{\textbf{E}}_{\pm}^{rd} = \frac{\omega\eta n_{mol}\alpha}{2r^2k}G\left(3\hat{u}_r \hat{u}_r-\mathds{I}\right)\cdot {\mathrm{\textbf{E}}}_{\pm}^{c}.
\end{equation}

The near field has properties  $\textrm{E}_+^{x}=\textrm{E}_-^{x}$, $\textrm{E}_+^{y}=\textrm{-E}_-^{y}$, and $\textrm{E}_+^{z}=\textrm{E}_-^{z}$. According to Eq. (\ref{edr}), $\mathrm{\textbf{E}}_{+}^{rd}$ and $\mathrm{\textbf{E}}_{-}^{rd}$  has same enantiomorphic relationship with $\mathrm{\textbf{E}}_{+}^{0}$ and $\mathrm{\textbf{E}}_{-}^{0}$. Thereby, the electric field in achiral nanostructure still has properties $\textrm{E}_+^{0x}+\textrm{E}_+^{rdx}=\textrm{E}_-^{0x}+\textrm{E}_-^{rdx}$, $\textrm{E}_+^{0y}+\textrm{E}_+^{rdy}=-\textrm{E}_-^{0y}-\textrm{E}_-^{rdy}$, and $\textrm{E}_+^{0z}+\textrm{E}_+^{rdz}=\textrm{E}_-^{z}+\textrm{E}_-^{rdz}$. This means field intensities in achiral nanostructure for RCP and LCP illumination is equal. Therefore, racemic mixture of chiral molecule cannot induce ICD in nanostructure.

The radiated electric field by pure chiral molecule is given in Eq. (\ref{ed}). Calculate the increments of
electric field strength $\triangle\mathrm{\textbf{E}}_{\pm}^{d}=\mathrm{\textbf{E}}_{\pm}^{d}-\mathrm{\textbf{E}}_{\pm}^{rd}$ which is generated by molecular chirality, we derive (\hyperlink{ex:einc}{Supporting information, Section \uppercase\expandafter{\romannumeral2}. B})
\begin{eqnarray}
\label{eddiff}
 \triangle\mathrm{\textbf{E}}_{\pm}^{d}=\mathrm{\textbf{E}}_{\pm}^{d}-\mathrm{\textbf{E}}_{\pm}^{rd} = \frac{-i\eta n_{mol}\kappa}{2r^2}G\left(3\hat{u}_r \hat{u}_r-\mathds{I}\right)\cdot {\mathrm{\textbf{B}}}_{\pm}^c.
\end{eqnarray}

If calculate the  $x$ component of electric field increments $\triangle\mathrm{{E}}_{\pm}^{dx}$ which is generated by dipole moment along $x$ direction, we derive
\begin{equation}\label{dx}
\triangle\mathrm{{E}}_{\pm}^{dx}=\frac{-i\eta n_{mol} \kappa}{16\pi r^3}{\mathrm{\text{B}}}_{\pm}^{cx}e^{-ikr}.
\end{equation}

In a similar manner, the $y$ component of electric field increments $\triangle\mathrm{{E}}_{\pm}^{dy}$ generated by dipole moment along $y$ direction is
\begin{equation}\label{dy}
\triangle\mathrm{{E}}_{\pm}^{dy}=\frac{-i\eta n_{mol} \kappa}{16\pi r^3}{\mathrm{\text{B}}}_{\pm}^{cy}e^{-ikr}.
\end{equation}

The $\triangle\mathrm{{E}}_{\pm}^{dx}$ has anti-enantiomorphic relation $\triangle\mathrm{E}_{+}^{dx}= -\triangle\mathrm{E}_{-}^{dx}$ while $\mathrm{{E}}_{\pm}^{dx}$ has enantiomorphic relation
$\mathrm{E}_+^{0x}=\mathrm{E}_-^{0x}$, and $\triangle\mathrm{{E}}_{\pm}^{dy}$ has enantiomorphic relation $\triangle\mathrm{{E}}_{+}^{dy}= \triangle\mathrm{{E}}_{-}^{dy}$ while $\mathrm{{E}}_{\pm}^{dy}$ has  anti-enantiomorphic relation
$\mathrm{E}_+^{0y}=\mathrm{-E}_-^{0y}$. Since the enantiomorphic relation of  $\textbf{E}^{0}$ and $\textbf{E}^{d}$ is different,  electric field strength in nanostructure generated by RCP and LCP incident light is not equal anymore, that is
$\textrm{E}_+^{0x}+\textrm{E}_+^{rdx}\neq\textrm{E}_-^{0x}+\textrm{E}_-^{rdx}$, $\textrm{E}_+^{0y}+\textrm{E}_+^{rdy}\neq-\textrm{E}_-^{0y}-\textrm{E}_-^{rdy}$, and $\textrm{E}_+^{0z}+\textrm{E}_+^{rdz}\neq\textrm{E}_-^{z}+\textrm{E}_-^{rdz}$. The nonequal electric field strengths leading to different absorption for RCP and LCP. This is the mechanism of ICD induction in chiral molecule-achiral nanostructure near field interaction.

\begin{figure}[b]
\centering
\includegraphics[height=5.9cm]{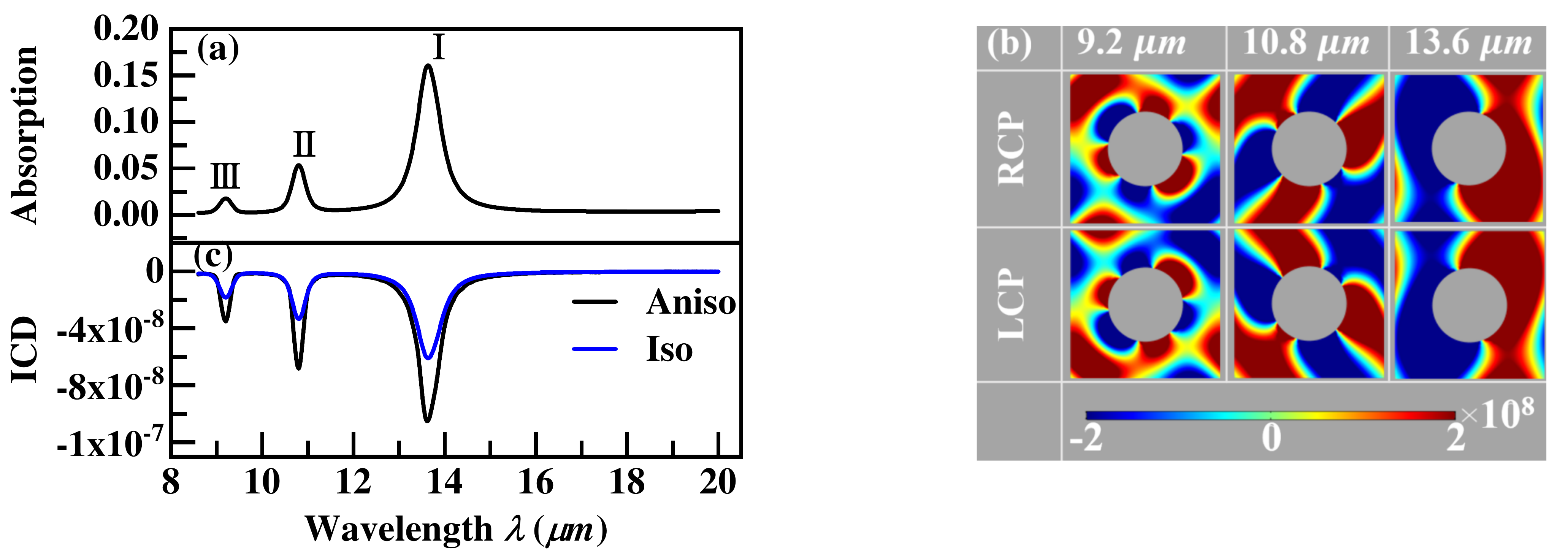}
\caption{(a) Absorption and ICD spectrum of GNM, (b) Charge distributions on surface of GNM.}
\label{fig22}
\end{figure}
\textbf{Comparison of the analytical results to the simulation results}. To prove the theory, we proposed the chiral molecule-GNM composed system, and investigated the ICD properties. Figure \ref{fig22}(a) depicts the absorption spectra of GNM. Three SPR modes (marked as \uppercase\expandafter{\romannumeral1}, \uppercase\expandafter{\romannumeral2}, and
\uppercase\expandafter{\romannumeral3}) in the absorption spectrum are located at wavelengths $13.6\,\mu m$, $10.8\,\mu m$ and $9.2\,\mu m$, respectively. Figure \ref{fig22}(b) depicts the
surface charge distribution on GNM. The red color corresponds to positive charge, the blue color corresponds to negative charge, and the charge distributions for RCP and LCP has enantiomorphic relation. Mode \uppercase\expandafter{\romannumeral1} is a
dipolar mode because the positive and negative charges are clearly separated in two regions. Mode \uppercase\expandafter{\romannumeral2} is a hexapolar mode, the positive and negative
charges are located in the six apexes of hexagon. Mode \uppercase\expandafter{\romannumeral3} is an octupolar mode composed of a hexapolar and dipolar mode. Figure. \ref{fig22}(c) shows the
ICD  spectra of GNM. Three  ICD dips are appear at three modes. Associated whit Eq. (\ref{cd}), we can conclude that the three ICD dips arise from electric field strength $\textbf{E}^{0}$ and magnetic flux density $\textbf{B}^{c}$ enhancement at three modes.  Since the field enhancement is largest at mode \uppercase\expandafter{\romannumeral1} while smallest at mode \uppercase\expandafter{\romannumeral3}, the  absorption and ICD  are also largest at mode \uppercase\expandafter{\romannumeral1} while smallest at mode \uppercase\expandafter{\romannumeral3}.
The
ICD induced by oriented molecules is larger than that in the case of the isotropic molecules and it obtains the maximum
when molecular orientation is parallel to $x$-axis. Therefore, molecular orientation is set as parallel to $x$- axis in the following.

\begin{figure}[h]
\centering
\includegraphics[height=5.9cm]{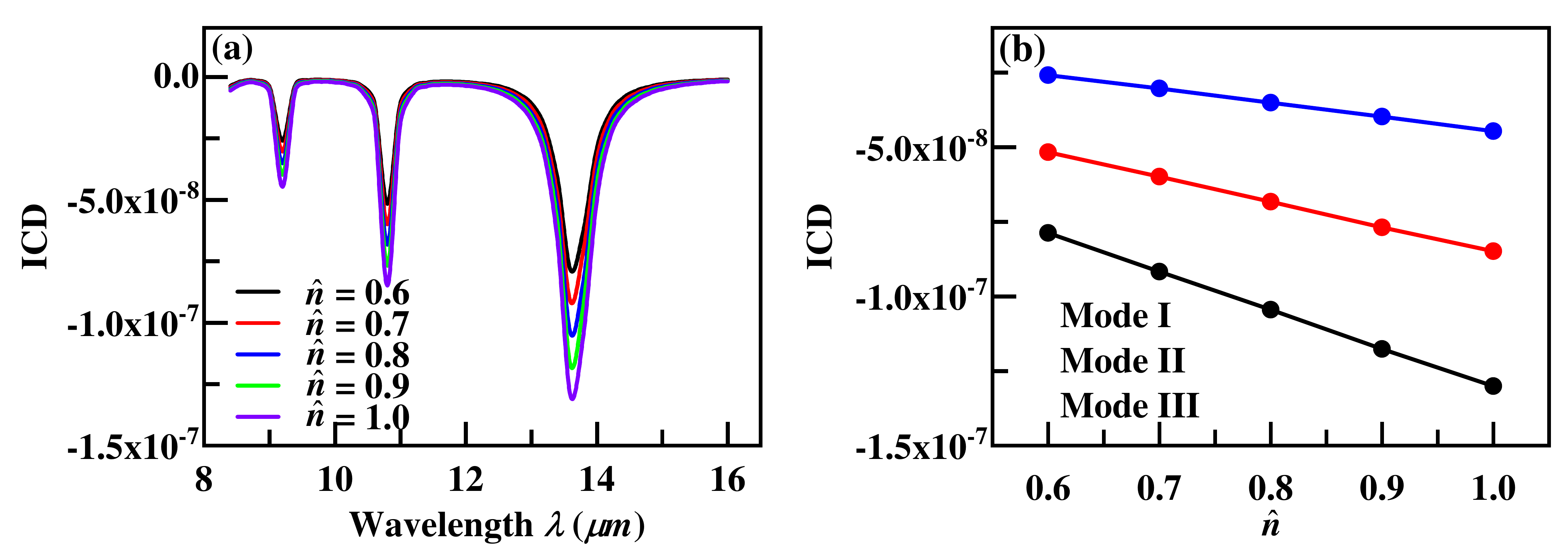}
\caption{(a) The  ICD spectra for different molecule densities, (b) Linear fitting (solid line)
of ICD and molecule densities at the three modes.}
\label{fig33}
\end{figure}

In order to prove the linear relation between molecular density and  ICD, the ICD for different molecular densities are calculated. For the sake of simplicity,
the density of molecule is normalized by $\hat{n}=\frac{n_{mol}}{6.25\times 10^{21}\,m^{-3}}$. Figure \ref{fig33}(a) presents the ICD spectra for different molecule number
densities $\hat{n}$. As expected, the ICD  peak value increases with the increase of $\hat{n}$. In order to compare with the analytical result, the linear fitting was
made at wavelengths of three modes. Figure \ref{fig33}(b) shows linear fitting of $\hat{n}-\textsc{ICD}$  relation at mode \uppercase\expandafter{\romannumeral1} (black line), mode
\uppercase\expandafter{\romannumeral2} (red line) and mode \uppercase\expandafter{\romannumeral3} (blue line). It can be seen that there are indeed linear relations between $\hat{n}$ and ICD, and the goodness of fitting is $0.99$. This is consistent with analytical result of Eq. (\ref{cd}).

\begin{figure}[h]
\centering
\includegraphics[height=5.9cm]{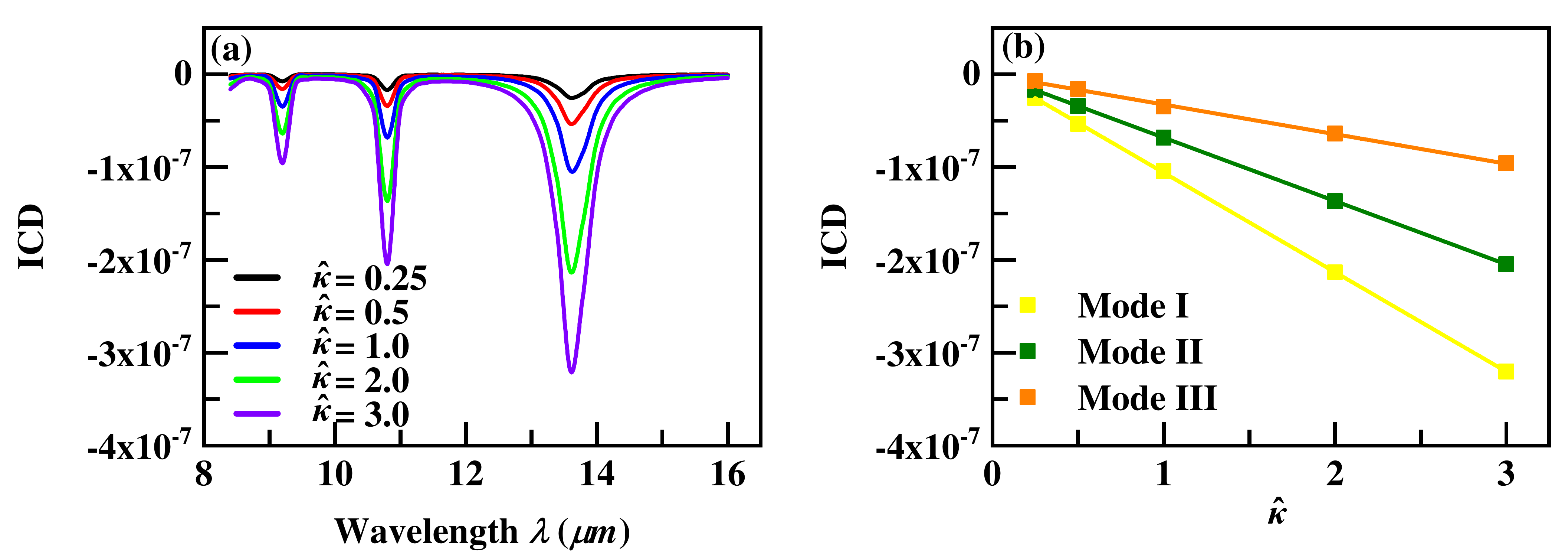}
\caption{(a) The ICD  spectra for different chirality parameter, (b) Linear fitting (sold line) of  ICD (Square) and chirality parameters at the three
modes.}
\label{fig44}
\end{figure}

Similarly, to prove the linear relation between molecular parameter $\kappa$ and ICD, the ICD for different molecular chirality parameters are calculated.
For the sake of simplicity, the chirality parameter is normalized by $\hat{\kappa}=\frac{\kappa_{mol}}{\kappa_{mol}^0}$. Here, $\kappa_{mol}^0$  is the initial molecular
chirality parameter which corresponds to $R_{rot} = 1.6\times10^{-11}m^3s^{-1}$  and $n_{mol} = 5\times10^{22}m^{-3}$. The ICD spectra with the change of
$\hat{\kappa}$  is given by Fig. \ref{fig44}(a). The ICD increases with increasing of $\hat{\kappa}$ at all three modes. Figure \ref{fig44}(b) shows the linear fitting at mode
\uppercase\expandafter{\romannumeral1} (yellow line), mode \uppercase\expandafter{\romannumeral2} (olive line) and mode \uppercase\expandafter{\romannumeral3} (orange line). The
fitting shows linear relationships between $\hat{\kappa}$  and ICD. Same as fitting of $\hat{n}-\textsc{ICD}$ , goodness of  fitting for $\hat{\kappa}-\textsc{ICD}$ relation
is 0.99. These results further prove proposed theory.

\begin{figure}[t]
\centering
\includegraphics[height=5.9cm]{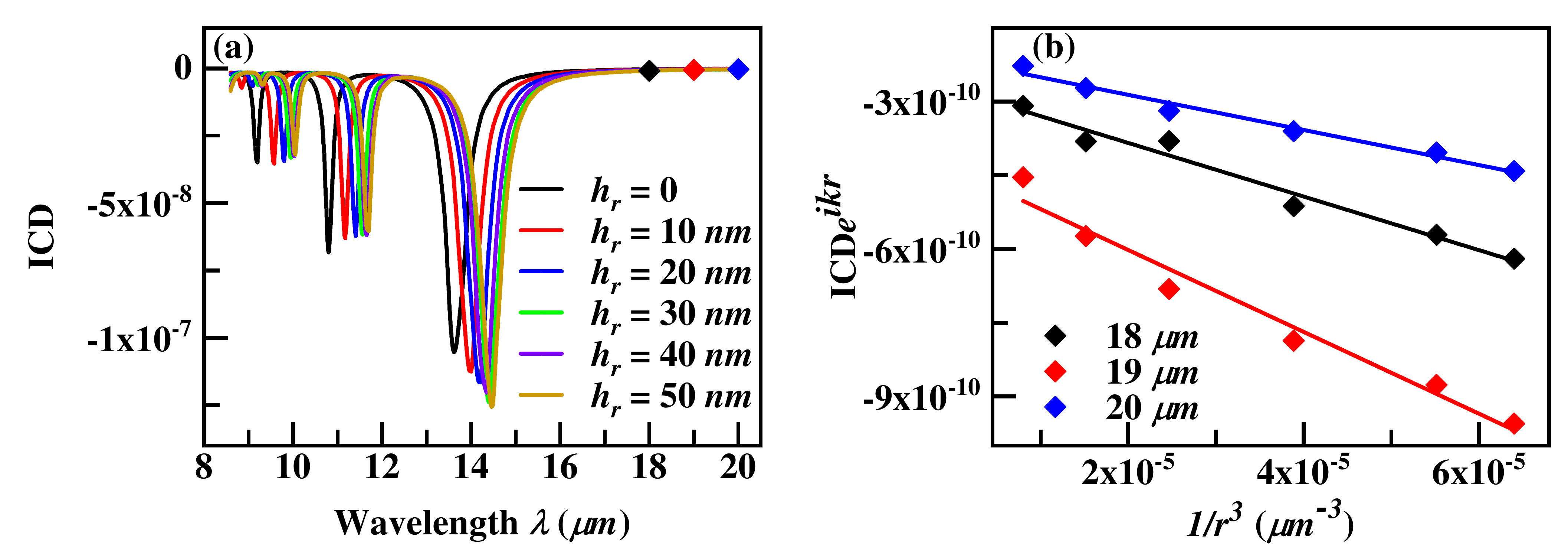}
\caption{(a) The ICD  spectra for different $h_r$, (b) The linear fitting of $\textsc{ICD}e^{ikr}-\frac{1}{r^3}$  relation.}
\label{fig55}
\end{figure}

In an effort to determine the relation between ICD  and molecule-nanostructure distance $r$, the distance $h_r$ from the GNM to bottom of molecular region is changed by a
step of $10\, nm$. As the GNM is moved from the bottom to the center of the chiral molecules layer, the ICD at mode \uppercase\expandafter{\romannumeral1} increases and
approaches the maximum when GNM reaches the center of chiral molecular region. And the  ICD at mode \uppercase\expandafter{\romannumeral2} and
\uppercase\expandafter{\romannumeral3} are decreased with increase of $h_r$.  According to Eq. (\ref{cd}), the relationship between $e^{ikr}\textsc{ICD}$ and
$\frac{1}{r^3}$ is linear, but electric field strength $\textbf{E}^{0}$ and magnetic flux density $\textbf{B}^{c}$ change with $h_r$ thereby affecting the ICD. Through comparing the $\textbf{E}^{0*}\cdot \textbf{B}^c$ and $\frac{\textbf{E}^{0*}\cdot \textbf{B}^c}{r^3}$ spectra (\hyperlink{ex:r}{Fig.} \ref{r}), we concluded that the influence of distance $\textbf{r}$ is obvious, and the $\frac{\textbf{E}^{0*}\cdot \textbf{B}^c}{r^3}$ spectra are more close to ICD spectra. To eliminate the effect of $\textbf{E}^{0}$ and $\textbf{B}^{c}$ changes and inconvenience for fitting due to different peak positions for different $h_r$, three off-resonance wavelengths $18\,{\mu}m$ , $19\,\mu m$  and $20\,{\mu}m$ are chosen to get liner fitting of
 $e^{ikr}\textsc{ICD}-\frac{1}{r^3}$  relation. Because effects of mode shape changes on $\textbf{E}^{0}$ and $\textbf{B}^{c}$ for different $h_r$ are negligible at the three
wavelengths.
To fitting liner relation of $e^{ikr}\textsc{ICD}$ and $\frac{1}{r^3}$  , the averaged distance between molecule layer to GNM  is calculated by $r=\frac{\sum_i(z_i-z_0)}{i}$, here, $z_0$ is the $z$ coordinate
of graphene and the  $z_i$ is the $z$ coordinate of chiral molecule layers. Figure. \ref{fig55}(b) shows the linear fitting at the three wavelengths. The fitting shows linear relations
between $e^{ikr}\textsc{ICD}$ and$\frac{1}{r^3}$, and the $e^{ikr}\textsc{ICD}-\frac{1}{r^3}$ relations are well consistent with the analytical results.

\begin{figure}[b]
\centering
\includegraphics[width=16.5cm]{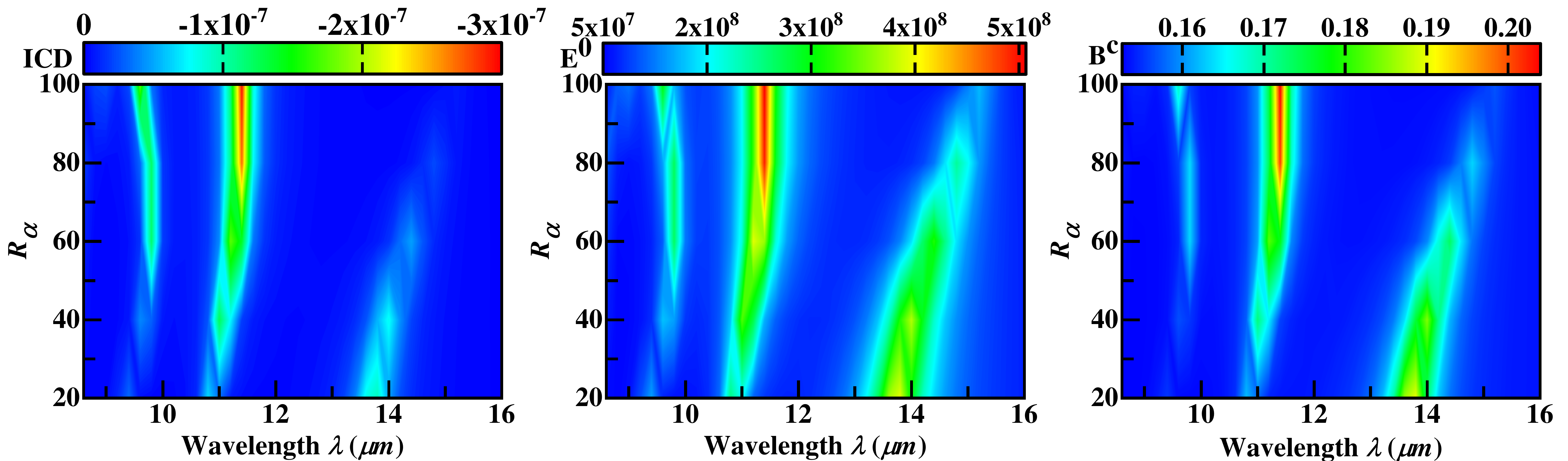}
\caption{The (a) ICD, (b) $\textsc{E}^{0}$ and (c) $\textsc{B}^{c}$  spectra for different $R_{\alpha}$.}
\label{fig66}
\end{figure}

For the purpose of investigating the effects of $\textbf{E}^{0}$ and $\textbf{B}^{c}$ on the ICD, we change the structural parameters to regulate
$\textbf{E}^{0}$ and $\textbf{B}^{c}$ (\hyperlink{ex:ebchange}{Supporting information, Section \uppercase\expandafter{\romannumeral2}. D}). A small additional hole is set at the center of every four adjacent original holes. And the radius of additional hole is $R_{\alpha}$  is
changed from $20\,nm$ to $100\,nm$ by a step of $20\,nm$. Figure \ref{fig66}(a) shows the ICD  spectra for different $R_{\alpha}$. After adding the hole, the
ICD  decreases at mode \uppercase\expandafter{\romannumeral1}, increases at mode \uppercase\expandafter{\romannumeral2} and mode
\uppercase\expandafter{\romannumeral3}. Figure \ref{fig66}(b) and (c) shows the averaged $\textbf{E}^{0}$ and  $\textbf{B}^{c}$ spectra for different  $R_{\alpha}$.
From figure \ref{fig66} it can be observed that the change of the $\textbf{E}^{0}$, $\textbf{B}^{c}$ and ICD are similar, the reason is that
ICD spectra is  regulated by the electric field $\textbf{E}^{0}$ and magnetic field $\textbf{B}^{c}$. At mode
\uppercase\expandafter{\romannumeral1}, ICD is increased by by $\textbf{E}^{0}$ and $\textbf{B}^{c}$, whereas at mode \uppercase\expandafter{\romannumeral2} and \uppercase\expandafter{\romannumeral3} the ICD gets decreased by $\textbf{E}^{0}$ and $\textbf{B}^{c}$ which agrees with Eq. (\ref{cd}).
\section{Conclusion}
Based on chiral molecule dipole approaching and chiral molecule-achiral nanostructure near field approximation, we present a theory for quantitative analysis of ICD from chiral molecule-achiral structure near field interaction. The theory clarifies the mechanism of ICD is that the dipole radiation of chiral molecules breaks the enantiomorphic relationship of the electric field strengths in the nanostructure. Furthermore, the ICD properties of graphene nanohole metasurface and chiral molecule coupled system are simulated to prove the
theory. The simulated results are consistent with the theoretical ones. There are linear relations between the ICD and molecular density, molecular chirality
parameter, and the ICD exponentially increases with the inverse cube of molecule-nanostructure distance. Finally, a smaller hole is added at the center of
four original holes to regulate the near field, and investigate the effect of electric field strength $\textbf{E}^{0}$ in nanostructure  and magnetic flux density $\textbf{B}^{c}$ in chiral
molecule region on the ICD. As expected, the ICD is  proportional to $\textbf{E}^{0}$ and $\textbf{B}^{c}$. Both, the
analytic and simulation results show two ways of enhancing the ICD. First, decreasing the molecule-nanostructure distance, and second, enhancing magnetic field in
molecular region or enhancing electric field inside nanostructure. These results are useful for analyzing experimental or calculated data, and beneficial for realizing supersensitive
detection of molecular chirality by plasmon enhancing technique.
\section{Acknowledgment}
We would like to thank Zhongyue Zhang  for useful discussion and valuable comments.  This work was supported by Fundamental Research Funds for the Central Universities (Grant No. 2018TS066).

\clearpage
\setcounter{equation}{0}

\setcounter{figure}{0}
\setcounter{section}{0}
\setcounter{subsection}{0}

\textit{Supporting Informations for}
\begin{center}
\large{\textbf{Quantitative Study on Circular Dichroism Induction from Achiral Nanostructure-Chiral Matter Near Field Interactions}}

\normalsize
\renewcommand{\thefootnote}{\fnsymbol{footnote}} Abuduwaili. Abudukelimu,$^1$ \footnote{Email:
Abduwelee@snnu.edu.cn}Ziyan Zhang,$^1$
 Tursunay. Yibibulla$^2$ and Muhammad. Ikram$^1$\\
\textit{1. School of Physics and Information Technology, Shaanxi Normal University, Xi'an, 710119, China}\\
\textit{2. School of Physics and Electronics, Central South University, Changsha, 410083, P. R. China}
\end{center}
\section{Circular dichroism induction by dipole radiated electric field}

Circular dichroism (CD) is absorption difference of right circularly polarized (RCP) and left circularly polarized (LCP) light, and it is a characteristic property of chiral media. The essential difference between chiral and achiral media lies in the constitutive relations. The constitutive relation for a chiral medium is given by \cite{SAHSihvola}
\renewcommand\theequation{S\arabic{equation}}
\begin{equation}
\begin{array}
{l}\textbf{D}=\varepsilon_0 \varepsilon_{c}{{\textbf{E}}}-\frac{i\kappa}{c}{\mathrm{\textbf{H}}} \\
\textbf{B}=\mu_0 \mu_{c}{{\textbf{H}}}+\frac{i\kappa}{c} {\mathrm{\textbf{E}}}
\end{array}
\end{equation}

where, $\varepsilon_{c}$, $\mu_{c}$ are respectively the relative permittivity and  magnetic permeability of the chiral molecule, $\kappa$ is the molecular chirality parameter, $\varepsilon_0$ and $ \mu_0$ are the vacuum permittivity and magnetic permeability, respectively.

Generally, the absorption of a nanostructure is given by\cite{SAOGovorov2012}
\renewcommand\theequation{S\arabic{equation}}
\begin{equation}
\label{Sabsorption1}
  {A}_{\pm}^{ns}=\frac{\omega}{2}\int_{V_{ns}}\left[\mathrm{Im}\left(\varepsilon_{ns}\right)\left({\mathrm{\textbf{E}}}_{\pm}^{0}\cdot{\mathrm{\textbf{E}}}_{\pm}^{0*}\right)+ \mathrm{Im}\left(\mu_{ns}\right)\left({\mathrm{\textbf{H}}}_{\pm}^{0}\cdot{\mathrm{\textbf{H}}}_{\pm}^{0*}\right)\right]dV,
\end{equation}
where, $\omega$  is angular frequency of CPL, $\varepsilon_{ns}$ and $\mu_{ns}$ are relative permittivity and permeability of nanostructure, ${\mathrm{\textbf{E}}}_{\pm}^{0}$ and ${\mathrm{\textbf{H}}}_{\pm}^{0}$ are electric and magnetic field strengths. The magnetic susceptibility of plasmonic nanostructure is very small, therefore magnetic loss can be neglected. In presence of chiral molecule, ${\mathrm{\textbf{E}}}_{\pm}^{0}$ will be affected by the electric field radiated by dipole moment of chiral molecule. Then Eq. (\ref{Sabsorption1}) can be written \cite{SBHan}
\renewcommand\theequation{S\arabic{equation}}
\begin{equation}\label{Sabsorption2}
  {A}_{\pm}^{ns}=\frac{\omega}{2}\mathrm{Im}\left(\varepsilon_{ns}\right)\int_{V_{ns}}\left({\mathrm{\textbf{E}}}_{\pm}^{0}+ {\mathrm{\textbf{E}}}_{\pm}^{d}\right)^{\ast}\cdot\left({\mathrm{\textbf{E}}}_{\pm}^{0}+ {\mathrm{\textbf{E}}}_{\pm}^{d}\right)dV,
\end{equation}
where, ${\mathrm{\textbf{E}}}_{\pm}^{d}$ is dipole radiated electric field  which can be written in terms of Green's function.
For time harmonic electromagnetic fields, the Green functions are dyadic functions, because the source is vector. Thus, the dipole radiated electric field can be written in terms of Green dyadic integral form
\renewcommand\theequation{S\arabic{equation}}
\begin{equation}
\label{Sedg}
\mathrm{\textbf{E}}^{d}\left(r\right)=\int_{V'}\left[\overline{\overline{G}}_{ee}\left(\textbf{r}-\textbf{r}' \right)\qquad \overline{\overline{G}}_{em}\left(\textbf{r}-\textbf{r}' \right)\right]\left(
  \begin{array}{ccc}
    \textbf{J} \\
    \textbf{M} \\
  \end{array}
\right)
\end{equation}
where the $\textbf{J}$ and $\textbf{M}$ electric and magnetic current. The Green functions satisfy the Maxwell equations. Using chiral constitutive relation to solve the Maxwell equations we derive Green functions for chiral medium\cite{SIVLindell}
\renewcommand\theequation{S\arabic{equation}}
\begin{eqnarray}
\label{SGreenee}
\overline{\overline{G}}_{ee}\left(r\right)&=&\overline{\overline{G}}_{ee+}\left(r\right)+\overline{\overline{G}}_{ee-}\left(r\right)\nonumber\\
\overline{\overline{G}}_{ee\pm}\left(r\right)& =&-\frac{i\eta}{2}\left[k_\pm\mathrm{G}_{\pm}\left(r\right)\mathds{I}\pm\nabla\mathrm{G}_{\pm}\left(r\right)\times \mathds{I}+\nabla\nabla\left(\frac{\mathrm{G}_{\pm}\left(r\right)}{k_\pm}\right)\right],\\
\overline{\overline{G}}_{em}\left(r\right)&=&\overline{\overline{G}}_{em+}\left(r\right)+\overline{\overline{G}}_{em-}\left(r\right)\nonumber\\
\overline{\overline{G}}_{em\pm}\left(r\right)&=&-\frac{1}{2}\left[\pm k_\pm\mathrm{G}_{\pm}\left(r\right)\mathds{I}+\nabla\mathrm{G}_{\pm}\left(r\right)\times \mathds{I}\pm\nabla\nabla\left(\frac{\mathrm{G}_{\pm}\left(r\right)}{k_\pm}\right)\right]\nonumber
\end{eqnarray}
where $\mathrm{G}_{\pm}\left(r\right)=\frac{e^{-ik_{\pm}r}}{4\pi r}$, $k_{\pm}=k\left(1\pm\frac{\kappa}{n_c}\right)$ and $n_c$ is refractive index of chiral medium. Using the approximation $ k_{\pm}r\ll 1$ for near field calculation, and calculating the gradient terms in Eq. (\ref{SGreenee}) becomes
\renewcommand\theequation{S\arabic{equation}}
\begin{eqnarray}
  \nabla \mathrm{G}_{\pm}\left(r\right) &=& \left(-ik_\pm+\frac{1}{r}\right)\mathrm{G}_{\pm}\left(r\right)\hat{u}_r\approx \frac{1}{r}\mathrm{G}_{\pm}\left(r\right)\hat{u}_r, \nonumber\\
  \nabla\nabla\mathrm{G}_{\pm}\left(r\right) &=& \left[\left(ik_\pm+\frac{1}{r}\right)^2+\frac{1}{r^2}\right]\mathrm{G}_{\pm}\left(r\right)\hat{u}_r \hat{u}_r-\left(ik_\pm+\frac{1}{r}\right)\frac{1}{r}\left(\mathds{I}-\hat{u}_r \hat{u}_r\right)\mathrm{G}_{\pm}\left(r\right)\\
  &\approx&\frac{3}{r^2}\frac{\mathrm{G}_{\pm}\left(r\right)}{k_{\pm}}\hat{u}_r\hat{u}_r-\frac{1}{r^2}\frac{\mathrm{G}_{\pm}\left(r\right)}{k_{\pm}}\mathds{I}.\nonumber
\end{eqnarray}
where $\hat{u}_r$ is unit vector in $\textbf{r}$ direction, $\hat{u}_r\hat{u}_r$ is dyadic, and can be written as,
\renewcommand\theequation{S\arabic{equation}}
\begin{eqnarray}
\hat{u}_r =\frac{1}{r}\left(\begin{array}{ccc}
    x \\
    y \\
    z \\
  \end{array}
  \right), \qquad
  \hat{u}_r\hat{u}_r = \frac{1}{r^2}\left(\begin{array}{ccc}
    x^2 &xy &xz \\
    yx  &y^2 &yz \\
    zx &zy &z^2 \\
  \end{array} \right).\nonumber
  \end{eqnarray}
  Thus the dyadic Green functions becomes
  \renewcommand\theequation{S\arabic{equation}}
  \begin{eqnarray}
    \overline{\overline{G}}_{ee+}\left(r\right) &=&\frac{-i\eta}{2} \left[k_+G_+\left(r\right)-\frac{1}{r}G_+\left(r\right)\hat{u}_r\times\mathds{I}+\frac{3}{r^2}\frac{G_+\left(r\right)}{k_+}\hat{u}_r\hat{u}_r-\frac{1}{r^2}\frac{G_+\left(r\right)}{k_+}\mathds{I} \right],\nonumber \\
   \overline{\overline{G}}_{ee-}\left(r\right) &=&\frac{-i\eta}{2} \left[k_-G_-\left(r\right)+\frac{1}{r}G_-\left(r\right)\hat{u}_r\times\mathds{I}+\frac{3}{r^2}\frac{G_-\left(r\right)}{k_-}\hat{u}_r\hat{u}_r-\frac{1}{r^2}\frac{G_-\left(r\right)}{k_-}\mathds{I} \right].
  \end{eqnarray}
  Implementing $ k_{\pm}r\ll 1$, we derive
  \renewcommand\theequation{S\arabic{equation}}
  \begin{eqnarray}
    \overline{\overline{G}}_{ee+}\left(r\right) &=&\frac{-i\eta}{2r^2k} \left[{3}G_+\left(r\right)\hat{u}_r\hat{u}_r-G_+\left(r\right)\mathds{I} \right],\nonumber \\
    \overline{\overline{G}}_{ee-}\left(r\right) &=&\frac{-i\eta}{2r^2k} \left[{3}G_-\left(r\right)\hat{u}_r\hat{u}_r-G_-\left(r\right)\mathds{I} \right],
  \end{eqnarray}
where $k_\pm=k(1\pm\frac{\kappa}{1.77})$. Since $\kappa$ is of the order of $10^{-5}$, we can approximate $k_\pm=k$. Finally, we derive the dyadic Green functions for near field
  \renewcommand\theequation{S\arabic{equation}}
  \begin{eqnarray}
  \label{gee}
   \overline{\overline{G}}_{ee}\left(r\right) &=& \frac{-i\eta}{2r^2k}\left[G_+\left(r\right)+G_-\left(r\right)\right] \left({3}\hat{u}_r\hat{u}_r-\mathds{I} \right), \nonumber\\
&=& \frac{-i\eta}{2r^2k}\frac{1}{4\pi r}e^{-ikr}\left(e^{-ik\frac{\kappa}{n_c}}+e^{ik\frac{\kappa}{n_c}}\right)\left({3}\hat{u}_r\hat{u}_r-\mathds{I} \right), \\
&=& \frac{-i\eta}{2r^2k}\frac{1}{2\pi r} e^{-ikr}cosh\left(-ikr\frac{\kappa}{n_c}\right)\left({3}\hat{u}_r\hat{u}_r-\mathds{I} \right), \nonumber
\end{eqnarray}
In the same method we derive
\begin{eqnarray}
\label{gem}
      \overline{\overline{G}}_{em}\left(r\right) &=& \frac{-1}{2r^2k}\left[G_-\left(r\right)+G_-\left(r\right)\right]\left({3}\hat{u}_r\hat{u}_r-\mathds{I} \right), \nonumber\\
&=& \frac{-1}{2r^2k}\frac{1}{4\pi r}e^{-ikr}\left(e^{-ik\frac{\kappa}{n_c}}-e^{ik\frac{\kappa}{n_c}}\right)\left({3}\hat{u}_r\hat{u}_r-\mathds{I} \right), \\
&=& \frac{-1}{2r^2k}\frac{1}{2\pi r} e^{-ikr}sinh\left(-ikr\frac{\kappa}{n_c}\right)\left({3}\hat{u}_r\hat{u}_r-\mathds{I} \right)\approx 0.\nonumber
\end{eqnarray}

From Eqs. (\ref{Sedg}), (\ref{gee}) and (\ref{gem}), we derive
\begin{equation}
  \label{SEDP}
  \textbf{E}_\pm^d= \int_{V'}\overline{\overline{G}}_{ee}\cdot\textbf{J} dV'.
\end{equation}

The electric current can be calculated by $\textbf{J}_\pm=\frac{\partial}{\partial t}\textbf{P}_\pm=i\omega \textbf{P}_\pm$. And the polarization strength of chiral molecule is given by \cite{SYCTang}
  \renewcommand\theequation{S\arabic{equation}}
  \begin{equation}
  \label{Sp}
  \textbf{P}_\pm=n_{mol}\alpha\textbf{E}_\pm^c-n_{mol}\frac{i\kappa}{c}\textbf{B}_\pm^c,
\end{equation}
where $n_{mol}$ is the number density of molecule.  Thus, the  radiated electric field by chiral molecular dipole is
  \renewcommand\theequation{S\arabic{equation}}
  \begin{equation}
  \label{Sedgr}
  \textbf{E}_\pm^d=i\omega n_{mol}\alpha\overline{\overline{G}}\cdot\textbf{E}_\pm^c+\omega n_{mol}\frac{\kappa}{c}\overline{\overline{G}}\cdot\textbf{B}_\pm^c.
\end{equation}

The induced $\textsc{ICD}$ in the nanostucture is
  \renewcommand\theequation{S\arabic{equation}}
  \begin{eqnarray}
  {\textrm{ICD}}&=&\frac{\omega}{2}\mathrm{Im}(\varepsilon_{ns})\int_{V_{ns}}({\mathrm{\textbf{E}}}_{+}^{0}+{\mathrm{\textbf{E}}}_{+}^{d})^{\ast}\cdot({\mathrm{\textbf{E}}}_{+}^{0}+ {\mathrm{\textbf{E}}}_{+}^{d})dV-\frac{\omega}{2}\mathrm{Im}(\varepsilon_{ns})\int_{V_{ns}}({\mathrm{\textbf{E}}}_{-}^{0}+ {\mathrm{\textbf{E}}}_{-}^{d})^{\ast}\cdot({\mathrm{\textbf{E}}}_{-}^{0}+ {\mathrm{\textbf{E}}}_{-}^{d})dV\nonumber\\
  &=&\frac{\omega}{2}\mathrm{Im}(\varepsilon_{ns})\int_{V_{ns}}\left[ \left({\mathrm{\textbf{E}}}_{+}^{0}+{\mathrm{\textbf{E}}}_{+}^{d}\right)^{\ast}\cdot\left({\mathrm{\textbf{E}}}_{+}^{0}+ {\mathrm{\textbf{E}}}_{+}^{d}\right)-\left({\mathrm{\textbf{E}}}_{-}^{0}+ {\mathrm{\textbf{E}}}_{-}^{d}\right)^{\ast}\cdot({\mathrm{\textbf{E}}}_{-}^{0}+ {\mathrm{\textbf{E}}}_{-}^{d})\right]dV.
\end{eqnarray}

The dipole radiated electric field $\textbf{E}^d$ is much smaller than original near field in nanostructure. Therefore, after neglecting the second order term $\textbf{E}^d\cdot\textbf{E}^{d\ast}$ we derive
  \renewcommand\theequation{S\arabic{equation}}
  \begin{equation}
 {\textrm{ICD}}=\omega\mathrm{Im}(\varepsilon_{ns})\int_{V_{ns}}\textrm{Re}\left( \textbf{E}_+^0\cdot\textbf{E}_+^{d\ast}-\textbf{E}_-^0\cdot\textbf{E}_-^{d\ast}\right)dV.
\end{equation}

Replace $\textbf{E}^d$ to calculate $\textbf{E}_+^0\cdot\textbf{E}_-^{d\ast}$
  \renewcommand\theequation{S\arabic{equation}}
  \begin{eqnarray}
  \label{Semed}
 &&\left(\textbf{E}_\pm^{0*}\cdot\textbf{E}_\pm^d\right)=\textbf{E}_\pm^{0*}\cdot\left[\frac{\omega\eta n_{mol}\alpha}{2r^2k}G\left(3\hat{u}_r
  \hat{u}_r-\mathds{I}\right){\mathrm{\textbf{E}}}_{\pm}^c
  +\frac{-i\omega\eta n_{mol}}{2r^2k}\frac{\kappa}{c}G\left(3\hat{u}_r
  \hat{u}_r-\mathds{I}\right){\mathrm{\textbf{B}}}_{\pm}^c \right] \\
  &&=\frac{3\omega\eta n_{mol}\alpha}{2r^2k}\frac{1}{r^2}G\textrm{E}_\pm^{0x*}\left(x^2 \textrm{E}_\pm^{cx}+xy\textrm{E}_\pm^{cy}+xz\textrm{E}_\pm^{cz}\right)+ \frac{3\omega\eta n_{mol}\alpha}{2r^2k}\frac{1}{r^2}G\textrm{E}_\pm^{0y*}\left(yx \textrm{E}_\pm^{cx}+y^2\textrm{E}_\pm^{cy}+yz\textrm{E}_\pm^{cz}\right)\nonumber\\
 &&+ \frac{3\omega\eta n_{mol}\alpha}{2r^2k}\frac{1}{r^2}G\textrm{E}_\pm^{0z*}\left(zx \textrm{E}_\pm^{cx}+zy\textrm{E}_\pm^{cy}+z^2\textrm{E}_\pm^{cz}\right)
 -\frac{\omega\eta
  n_{mol}\alpha}{2r^2k}G\left(\textrm{E}_\pm^{0z*}\textrm{E}_\pm^{cx}+\textrm{E}_\pm^{0z*}\textrm{E}_\pm^{cy}+\textrm{E}_\pm^{0z*}\textrm{E}_\pm^{cz}\right)\nonumber\\
 &\;&-\frac{3i\omega\eta n_{mol}\kappa}{2r^2kc}\frac{1}{r^2}G\textrm{E}_\pm^{0x*}\left(x^2 \textrm{B}_\pm^{cx}+xy\textrm{B}_\pm^{cy}+xz\textrm{B}_\pm^{cz}\right)
  -\frac{3i\omega\eta n_{mol}\kappa}{2r^2kc}\frac{1}{r^2}G\textrm{E}_\pm^{0y*}\left(yx \textrm{B}_\pm^{cx}+y^2\textrm{B}_\pm^{cy}+zy\textrm{B}_\pm^{cz}\right)\nonumber\\
   &\;&-\frac{3i\omega\eta n_{mol}\kappa}{2r^2kc}\frac{1}{r^2}G\textrm{E}_\pm^{0z*}\left(zx \textrm{B}_\pm^{cx}+zy\textrm{B}_\pm^{cy}+z^2\textrm{B}_\pm^{cz}\right)
    +\frac{i\omega\eta
 n_{mol\kappa}}{2r^2kc}G\left(\textrm{E}_\pm^{0x*}\textrm{B}_\pm^{cx}+\textrm{E}_\pm^{0x*}\textrm{B}_\pm^{cy}+\textrm{E}_\pm^{0x*}\textrm{B}_\pm^{cz}\right),\nonumber
  \end{eqnarray}
 where $G=G_+\left(r\right)+G_-\left(r\right)$.

If we suppose that the incident light propagates along $z$ direction and the achiral naonostructure is symmetric to $y$ or $x$-axis (any achiral structure has a symmetry axis), then near field in the $xy$-plane has properties $\textrm{E}_+^x(x,y)=\textrm{E}_-^x(x,-y)$, $\textrm{E}_+^y(x,y)=\textrm{-E}_-^y(x,-y)$, $\textrm{E}_+^z(x,y)=\textrm{E}_-^z(x,-y)$, $\textrm{B}_+^x(x,y)=\textrm{-B}_-^x(x,-y)$, $\textrm{B}_+^y(x,y)=\textrm{B}_-^y(x,-y)$, $\textrm{B}_+^z(x,y)=\textrm{-B}_-^z(x,-y)$. Applying coordinate transformations $y\rightarrow-y$ or $x\rightarrow-x$ (depending on the axis of symmetry) to $\frac{1}{r^2}(x-x')^2 \textrm{E}_+^{0x'\ast}(x',y')\textrm{E}_+^{cx}(x,y)$ , we derive
 \renewcommand\theequation{S\arabic{equation}}
\begin{equation}
\label{Strans}
 \frac{1}{r^2}(x-x')^2\textrm{E}_+^{0x'\ast}(x',-y')\textrm{E}_+^{cx}(x,-y)=\frac{1}{r^2}(x-x')^2 \textrm{E}_-^{0x'\ast}(x',y')\textrm{E}_-^{cx}(x,y).
\end{equation}

Eq. (\ref{Strans}) means the term $\frac{1}{r^2}(x-x')^2 \textrm{E}^{0x'\ast}(x',y')\textrm{E}^{cx}(x,y)$ is equal for RCP and LCP illumination. Using the same coordinate transformations to other terms , we finally derive
  \renewcommand\theequation{S\arabic{equation}}
  \begin{eqnarray}
  \label{SSS15}
\textbf{E}_+^0\cdot\textbf{E}_+^{d\ast}-\textbf{E}_-^0\cdot\textbf{E}_-^{d\ast} &=&\frac{-6i\omega\eta n_{mol}}{r^2k}\frac{\kappa}{c}G\frac{1}{r^2}\left(x^2\textrm{E}_+^{0x\ast} \textrm{B}_+^{cx}+y^2\textrm{E}_+^{0y\ast} \textrm{B}_+^{cy}+z^2\textrm{E}_+^{0z\ast}\textrm{B}_\pm^{cz}\right)\\
  &+& \frac{2i\omega\eta n_{mol}}{r^2k}\frac{\kappa}{c}G\left(\textrm{E}_+^{0x\ast}\textrm{B}_+^{cx}+\textrm{E}_+^{0y\ast}\textrm{B}_+^{cy}+\textrm{E}_+^{0z\ast}\textrm{B}_+^{cz}\right).\nonumber
\end{eqnarray}

Eq. (\ref{SSS15}) is general form of field intensity difference generated by chiral molecular dipole. The $z$ component of near field is much smaller than $x$ and $y$ components, neglecting the $z$ component we derive
  \renewcommand\theequation{S\arabic{equation}}
  \begin{eqnarray}
  \label{Semed}
 \textbf{E}_+^0\cdot\textbf{E}_+^{d\ast}-\textbf{E}_-^0\cdot\textbf{E}_-^{d\ast} &=&\frac{-6i\omega\eta n_{mol}}{r^2k}\frac{\kappa}{c}G\frac{1}{r^2}\left(x^2\textrm{E}_+^{0x\ast} \textrm{B}_+^{cx}+y^2\textrm{E}_+^{0y\ast} \textrm{B}_+^{cy}\right)\\
  &+& \frac{2i\omega\eta n_{mol}}{r^2k}\frac{\kappa}{c}G\left(\textrm{E}_+^{0x\ast}\textrm{B}_+^{cx}+\textrm{E}_+^{0y\ast}\textrm{B}_+^{cy}\right).\nonumber
\end{eqnarray}

For planner nanostructure (in our GNM-chiral molecule composed system, $x$ and $y$ are infinite while $z$ is up to $\textrm{100}\, nm$ ) the the distance $r$ can approximated as $r=(x^2+y^2)^{1/2}$, then Eq. (\ref{Semed}) becomes
\renewcommand\theequation{S\arabic{equation}}
\begin{eqnarray}
\label{Ssincos}
&& \textbf{E}_+^0\cdot\textbf{E}_+^{d\ast}-\textbf{E}_-^0\cdot\textbf{E}_-^{d\ast}=\\
   &&\frac{-6i\omega\eta n_{mol}\kappa}{r^2kc}G\left(cos^2{\alpha}\textrm{E}_+^{0x\ast} \textrm{B}_+^{cx}+sin^2{\alpha}\textrm{E}_+^{0y\ast} \textrm{B}_+^{cy}\right)+\frac{2i\omega\eta n_{mol}\kappa}{r^2kc}G\left(\textrm{E}_\pm^{0x\ast}\textrm{B}_+^{cx}+\textrm{E}_+^{0y\ast}\textrm{B}_+^{cy}\right).\nonumber
\end{eqnarray}
where $\alpha$ is angle between $r$ and $x$-axis. If the chiral molecule is uniformly arranged around the achiral structure, taking average values of $sin^2(\alpha)$ and $cos^2(\alpha)$, Eq. (\ref{Ssincos}) becomes
\renewcommand\theequation{S\arabic{equation}}
\begin{eqnarray}
\label{SSsincos}
&&\textbf{E}_+^0\cdot\textbf{E}_+^{d\ast}-\textbf{E}_-^0\cdot\textbf{E}_-^{d\ast}=-\frac{i\omega\eta n_{mol}}{r^2k}\frac{\kappa}{c}G\left(\textrm{E}_+^{0x\ast}\textrm{B}_+^{cx}+\textrm{E}_+^{0y\ast}\textrm{B}_+^{cy}\right)\nonumber\\
&=&-\frac{i\omega\eta n_{mol}}{r^2k}\frac{\kappa}{c}\frac{1}{4\pi r}e^{-ikr}\left(e^{-ik\frac{\kappa}{n_c}}+e^{ik\frac{\kappa}{n_c}}\right)\left(\textrm{E}_+^{0x\ast}\textrm{B}_+^{cx}+\textrm{E}_+^{0y\ast}\textrm{B}_+^{cy}\right)\nonumber\\
&=&-\frac{i\omega\eta n_{mol}}{r^2k}\frac{\kappa}{c}\frac{1}{2\pi r} e^{-ikr}cosh\left(-ikr\frac{\kappa}{n_c}\right)\left(\textrm{E}_+^{0x\ast}\textrm{B}_+^{cx}+\textrm{E}_+^{0y\ast}\textrm{B}_+^{cy}\right)\nonumber\\
&=&-\frac{i\eta n_{mol}}{2\pi r^3}\kappa e^{-ikr}\left(\textrm{E}_+^{0x\ast}\textrm{B}_+^{cx}+\textrm{E}_+^{0y\ast}\textrm{B}_+^{cy}\right),
\end{eqnarray}
where $cosh\left(-ikr\frac{\kappa}{n_c}\right)\approx 1$.

Finally, we derive the $\textrm{ICD}$ generated by a molecular dipole strength $\textbf{P}$ in the nanostructure
\renewcommand\theequation{S\arabic{equation}}
\begin{equation}
\label{Scd}
  \textsc{ICD}=\frac{\eta n_{mol}}{2\pi}\mathrm{Im}(\varepsilon_{ns})\int_{V_{ns}}\frac{1}{r^3}\textrm{ Re} \left[-i\kappa e^{-ikr}\left(\textrm{E}_+^{0x\ast} \textrm{B}_+^{cx}+\textrm{E}_+^{0y\ast} \textrm{B}_+^{cy}\right)\right]dV.
\end{equation}

\section{Analyzing and approximations}
\subsection{Enantiomorphic properties of near field around achiral structure}
 Suppose if the incident light propagates along $z$ direction, the near field in $xy$-plain has enantiomorphic relations $\textrm{E}_+^x(x,y)=\textrm{E}_-^x(x,-y)$,  $\textrm{E}_+^z(x,y)=\textrm{E}_-^z(x,-y)$, $\textrm{B}_+^y(x,y)=\textrm{B}_-^y(x,-y)$ and anti-elastomeric relation $\textrm{E}_+^y(x,y)=\textrm{-E}_-^y(x,-y)$, $\textrm{B}_+^x(x,y)=\textrm{-B}_-^x(x,-y)$,  $\textrm{B}_+^z(x,y)=\textrm{-B}_-^z(x,-y)$. \hyperlink{EX:Sebxy}{Figure} \ref{Sebxy} shows the electric field and magnetic flux distribution of GNM at mode \uppercase\expandafter{\romannumeral2}.
The average electric and magnetic field values satisfy $\textrm{E}_+^x=\textrm{E}_-^x$,
 $\textrm{E}_+^y=\textrm{-E}_-^y$, $\textrm{E}_+^z=\textrm{E}_-^z$, $\textrm{B}_+^x=\textrm{-B}_-^x$, $\textrm{B}_+^y=\textrm{B}_-^y$, $\textrm{B}_+^z=\textrm{-B}_-^z$, ${\textrm{E}_+^z}\ll{\textrm{E}_+^x}\,,\, {\textrm{E}_+^y}$ and  ${\textrm{B}_+^z}\ll{\textrm{B}_+^x}\,,\, {\textrm{B}_+^y}$ . \hyperlink{ex:SAVAEB}{Figure} \ref{SAVAEB} shows the average electric field and magnetic flux spectra of GNM.

\renewcommand\thefigure{S\arabic{figure}}
\begin{figure}[h]
\centering
\includegraphics[height=8.5cm]{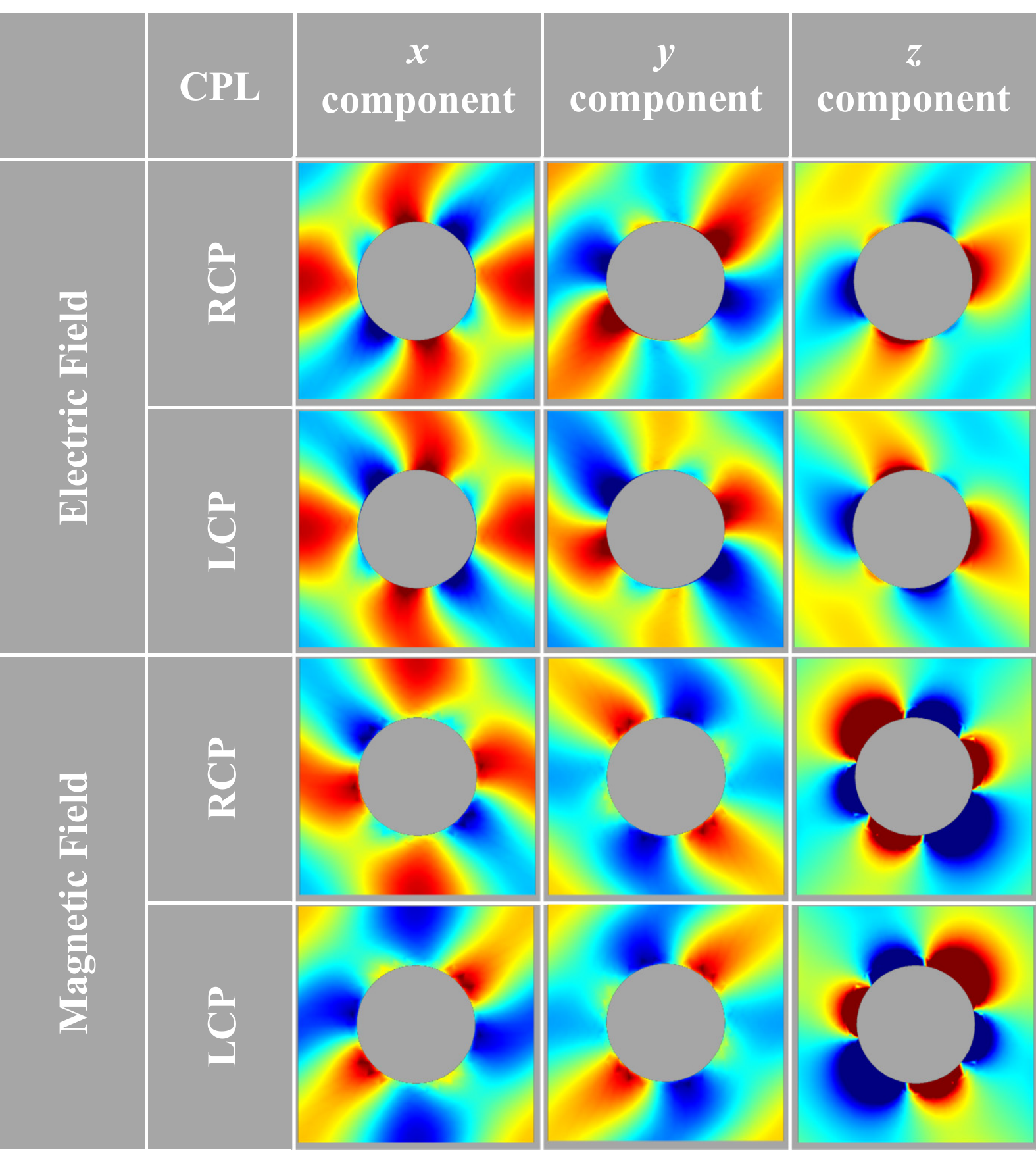}
\hypertarget{EX:Sebxy}{\caption{ Electric and magnetic field distribution at surface of GNM}}
\label{Sebxy}
\end{figure}

 \renewcommand\thefigure{S\arabic{figure}}
\begin{figure}[t]
\centering
\includegraphics[width=16.5cm]{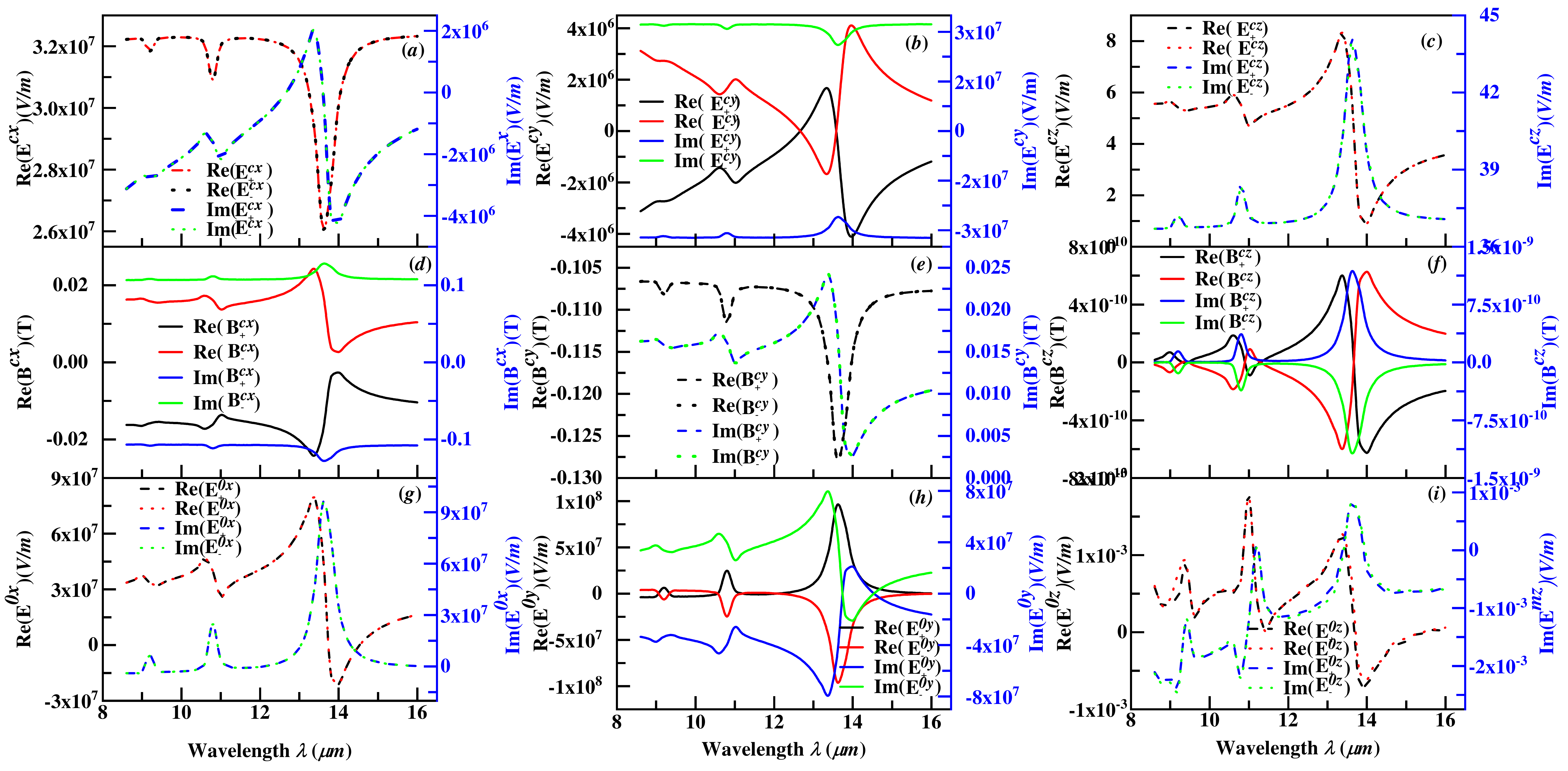}
\hypertarget{EX:SAVAEB}{\caption{Average $x$, $y$ and $z$ components of (a-c) electric field $\textbf{E}^c$, (d-f) magnetic field  $\textbf{B}^c$ and (g-i) electric field $\textbf{E}^{0}$.}}
\label{SAVAEB}
\end{figure}

\hypertarget{ex:einc}{\subsection{Enetiomeric relation breaking of electric fields in achiral nanostructure}} The racemic mixture of chiral molecule has no chirality and it does not induce $\textsc{ICD}$. The pure chiral molecule breaks the electric field symmetry in achiral structure and induces the $\textsc{ICD}$ signal. The averaged electric field in achiral nanostrcture and molecular region has properties  $\textrm{E}_+^{0x}=\textrm{E}_-^{0x}$, $\textrm{E}_+^{0y}=\textrm{-E}_-^{0y}$, $\textrm{E}_+^{0z}=\textrm{E}_-^{0z}$, and ${\mathrm{\textbf{E}}}_{\pm}^{c}$ satisfies the constitutive relations
\renewcommand\theequation{S\arabic{equation}}
\begin{eqnarray}
\label{erc}
\textbf{D}^{rc}&=&\varepsilon_0 \varepsilon_{c}\textbf{E}^{c}, \nonumber\\
\textbf{B}^{rc}&=&\mu_0 \mu_{c}\textbf{H}^{c}.\nonumber
\end{eqnarray}

The polarization strength of racemic mixture is given by
\renewcommand\theequation{S\arabic{equation}}
\begin{eqnarray}
\label{Prc}
\textbf{P}_{\pm}^r=n_{mol} \alpha\textbf{E}^{c}, \nonumber
\end{eqnarray}

The electric field radiated by the racemic mixture molecular dipole is
 \renewcommand\theequation{S\arabic{equation}}
 \begin{equation}\label{Sedr}
 \mathrm{\textbf{E}}_{\pm}^{rd} = \frac{\omega\eta n_{mol}\alpha}{2r^2k}G\left(3\hat{u}_r \hat{u}_r-\mathds{I}\right)\cdot {\mathrm{\textbf{E}}}_{\pm}^{c}.
\end{equation}

 According to Eq. (\ref{Sedr}) $\mathrm{\textbf{E}}_{+}^{rd}$ and $\mathrm{\textbf{E}}_{-}^{rd}$  has same enantiomorphic relationship with $\mathrm{\textbf{E}}_{+}^{0}$ and $\mathrm{\textbf{E}}_{-}^{0}$. The electric field in achiral nanostructure still hass properties $\textrm{E}_+^{0x}+\textrm{E}_+^{rdx}=\textrm{E}_-^{0x}+\textrm{E}_-^{rdx}$, $\textrm{E}_+^{0y}+\textrm{E}_+^{rdy}=-\textrm{E}_-^{0y}-\textrm{E}_-^{rdy}$, and $\textrm{E}_+^{0z}+\textrm{E}_+^{rdz}=\textrm{E}_-^{z}+\textrm{E}_-^{rdz}$. With the properties, field intensities in achiral nanostructure for RCP and LCP illumination is euqal. Therefore, racemic mixture of chiral molecule can not induce ICD in nanostructure.

If we calculate the difference of $\mathrm{\textbf{E}}_{\pm}^{rd}$ and $ \mathrm{\textbf{E}}_{\pm}^{d}$, we derive
\renewcommand\theequation{S\arabic{equation}}
\begin{equation}
\label{eddiff}
 \triangle\mathrm{\textbf{E}}_{\pm}^{d}=\mathrm{\textbf{E}}_{\pm}^{d}-\mathrm{\textbf{E}}_{\pm}^{rd} = \frac{-i\eta n_{mol}\kappa}{2r^2}G\left(3\hat{u}_r \hat{u}_r-\mathds{I}\right)\cdot {\mathrm{\textbf{B}}}_{\pm}^c.
\end{equation}

Calculating the electric field difference $\triangle\mathrm{{E}}_{\pm}^{dx}$ which is generated by dipole moment $n_{mol}\textrm{P}^{x}\hat{u}_x$ along $x$ axis, we derive
\renewcommand\theequation{S\arabic{equation}}
\begin{eqnarray}\label{dx}
\triangle\mathrm{{E}}_{\pm}^{dx}&=&\mathrm{{E}}_{\pm}^{d}-\mathrm{{E}}_{\pm}^{rd} = \frac{-i\eta n_{mol}\kappa}{2r^2}G\left(3\hat{u}_r \hat{u}_r-\mathds{I}\right)\cdot {\mathrm{{B}}}_{\pm}^{cx}\nonumber\\
&=&\frac{-i\eta n_{mol}\kappa}{2r^2}\frac{1}{4\pi r}e^{-ikr}e^{-ikr\frac{\kappa}{n_c}}\left(\frac{3x^2}{r^2}{{\text{B}}}_{\pm}^{cx}-{\mathrm{{B}}}_{\pm}^{cx}\right)\nonumber\\
&=&\frac{-i\eta n_{mol}\kappa}{8\pi r^3}e^{-ikr}\left(3{cos}^2\alpha{\mathrm{\text{B}}}_{\pm}^{cx}-{\mathrm{\text{B}}}_{\pm}^{cx}\right)
=\frac{-i\eta n_{mol} \kappa}{16\pi r^3}{\mathrm{\text{B}}}_{\pm}^{cx}e^{-ikr},
\end{eqnarray}
where $e^{-ikr\frac{\kappa}{n_c}}\approx 1$, and average value of ${cos}^2$ is 0.5.
\renewcommand\thefigure{S\arabic{figure}}
\begin{figure}[b]
\centering
\includegraphics[height=8cm]{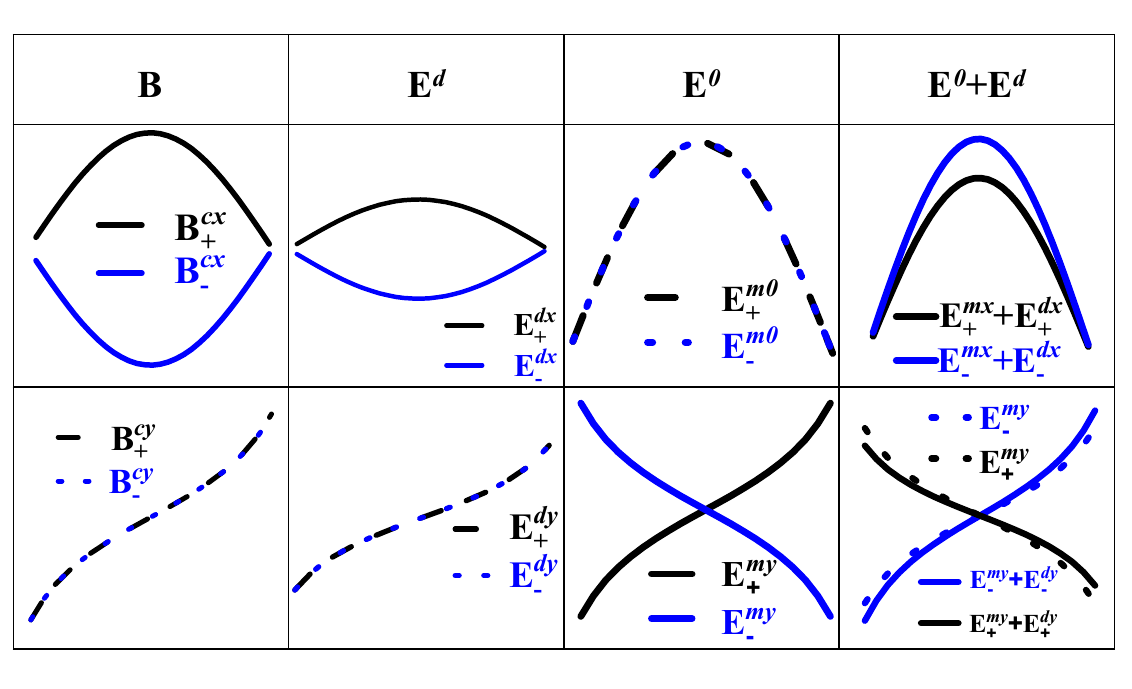}
\hypertarget{EX:Sfigemec}{\caption{ Schematic of electric field symmetry breaking in achiral nanostructure.}}
\label{figemec}
\end{figure}
 In a similar manner, the $y$ component of electric field increments $\triangle\mathrm{{E}}_{\pm}^{dy}$ generated by dipole moment along $y$ direction is
\begin{equation}\label{Sdy}
\triangle\mathrm{{E}}_{\pm}^{dy}=\frac{-i\eta n_{mol} \kappa}{16\pi r^3}{\mathrm{\text{B}}}_{\pm}^{cy}e^{-ikr}.
\end{equation}

The $\triangle\mathrm{{E}}_{\pm}^{dx}$ has anti-enantiomorphic relation $\triangle\mathrm{E}_{+}^{dx}= -\triangle\mathrm{E}_{-}^{dx}$ while $\mathrm{{E}}_{\pm}^{dx}$ has enantiomorphic relation
$\mathrm{E}_+^{0x}=\mathrm{E}_-^{0x}$, and $\triangle\mathrm{{E}}_{\pm}^{dy}$ has enantiomorphic relation $\triangle\mathrm{{E}}_{+}^{dy}= \triangle\mathrm{{E}}_{-}^{dy}$ while $\mathrm{{E}}_{\pm}^{dy}$ has  anti-enantiomorphic relation
$\mathrm{E}_+^{0y}=\mathrm{-E}_-^{0y}$. Since the enantiomeric relation of  $\textbf{E}^{0}$ and $\textbf{E}^{d}$ is different,  electric field strength in nanostructure generated by RCP and LCP incident light is not equal anymore, that is
$\textrm{E}_+^{0x}+\textrm{E}_+^{rdx}\neq\textrm{E}_-^{0x}+\textrm{E}_-^{rdx}$, $\textrm{E}_+^{0y}+\textrm{E}_+^{rdy}\neq-\textrm{E}_-^{0y}-\textrm{E}_-^{rdy}$, and $\textrm{E}_+^{0z}+\textrm{E}_+^{rdz}\neq\textrm{E}_-^{z}+\textrm{E}_-^{rdz}$. The nonequal electric field strengths leading to different absorption for RCP and LCP. This is the mechanism of ICD induction in chiral molecule-achiral nanostructure near field interaction. \hyperlink{EX:Sfigemec}{Figure} \ref{figemec} is schematic illustration the origin of electric field strength difference in achiral nanostructure. Generally, the increments in electric field by chiral molecule radiation are opposite to each other for RCP and LCP incident, if the electric field increments for RCP is positive, the electric field increments for LCP will be negative. \hyperlink{ex:Sfigde}{Figure} \ref{Sfigde} shows the electric field increments in GNM, the results are same as analytical result.

\renewcommand\thefigure{S\arabic{figure}}
\begin{figure}[h]
\centering
\includegraphics[height=6cm]{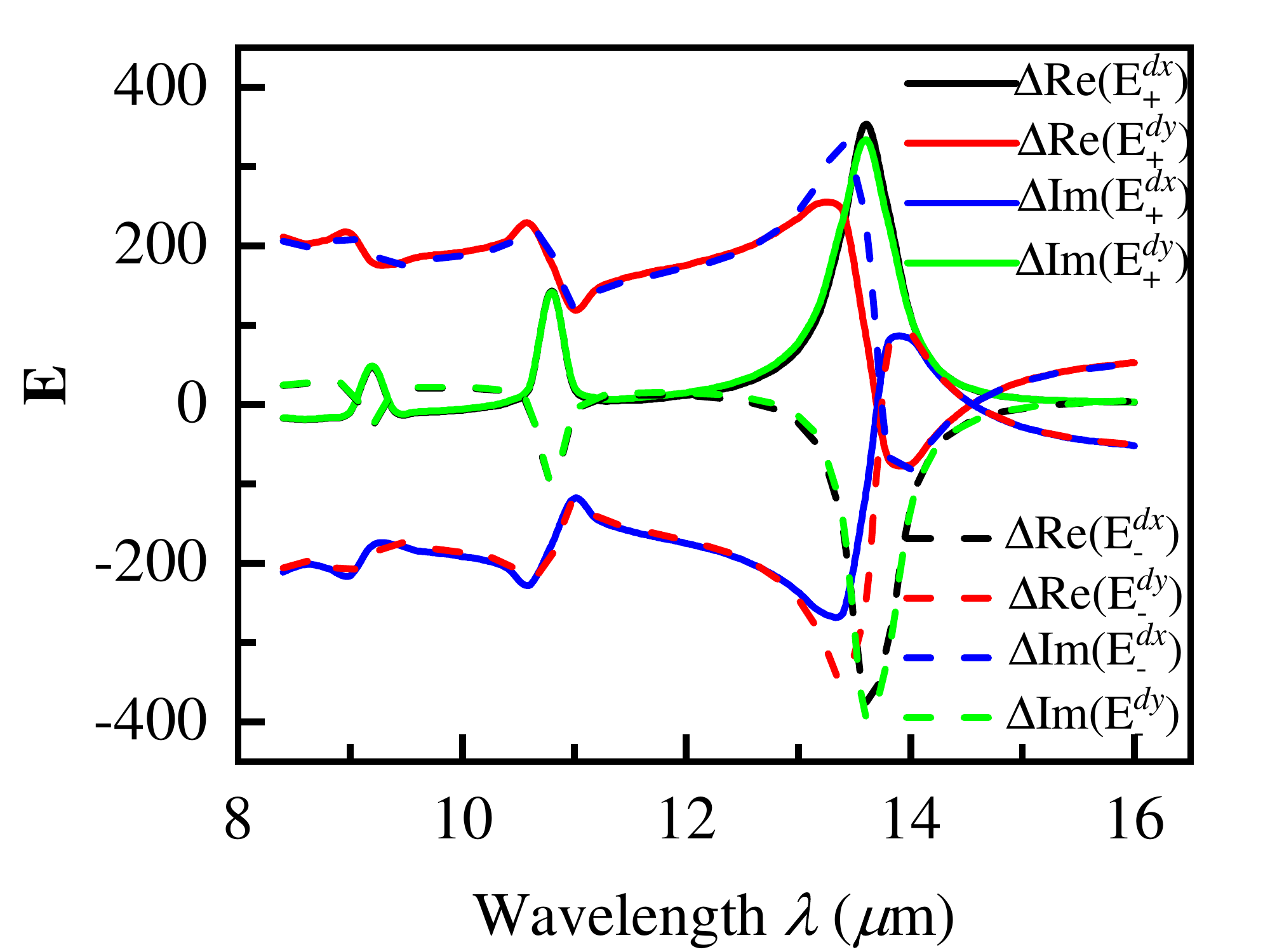}
\hypertarget{ex:Sfigde}{\caption{The electric field strength increments in GNM.}}
\label{Sfigde}
\end{figure}
\subsection{Influence of distance \textbf{r} to the ICD}
 According to Eq. (\ref{Scd}), there is a linear relationship between $\textsc{ICD}$ and
 $\frac{e^{-ikr}}{r^3}$. But the electric fields $\textrm{E}^{0}$ and magnetic fields $\textrm{B}^{c}$ change with $h_r$ thereby affecting the $\textsc{ICD}$. Thus we analyzed the $\textbf{E}^{0*}\cdot\textbf{B}^c$, $\frac{\textbf{E}^{0*}\cdot \textbf{B}^c}{r^3}$ and $\textsc{ICD}$.  \hyperlink{ex:r}{Figure} \ref{r} shows the $\textbf{E}^{0*}\cdot\textbf{B}^c$, $\frac{\textbf{E}^{0*}\cdot\textbf{B}^c}{r^3}$ and $\textsc{ICD}$ spectra. From $\textbf{E}^{0*}\cdot \textbf{B}^c$ and $\frac{\textbf{E}^{0*}\cdot\textbf{B}^c}{r^3}$ spectra it can be observed that the influence of distance $\textbf{r}$ is obvious, and the $\frac{\textbf{E}^{0*}\cdot \textbf{B}^c}{r^3}$ spectra are more close to $\textsc{ICD}$ spectra. Another inconvenience to do the linear fitting is that the peak wavelengths are different for different $h_r$. Therefore, for excluding the effect of $\textrm{E}^{0}$ and $\textrm{B}^{c}$, we chose the $ 18\, nm$, $19\,nm$ and $20\,nm$ for linear fitting. Due to these are off-resonance wavelength, electric and magnetic field strengths are almost same for different $h_r$.
\renewcommand\thefigure{S\arabic{figure}}
\begin{figure}[t]
\centering
\includegraphics[width=16.5cm]{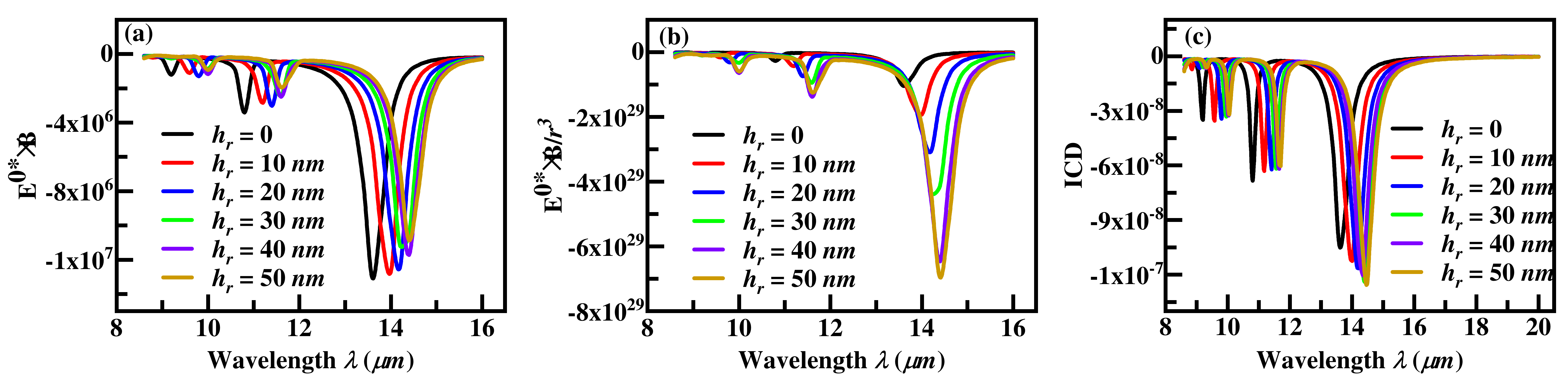}
\hypertarget{ex:r}{\caption{The spectra of (a) $\textbf{E}^{0*}\cdot\textbf{B}^c$, (b) $\textbf{E}^{0*}\cdot \textbf{B}^c/{r^3}$ and (c) $\textsc{ICD}$ for different $h_r$.}}
\label{r}
\end{figure}
\hypertarget{ex:ebchange}{\subsection{Effect of electric field in nanostructure and magnetic field in molecular region on ICD}}
The electric field strength ${{\mathbf{E}}^{0}}$ and magnetic field density ${{\mathbf{B}}^{c}}$ also affect the $\textrm{ICD}$. For the purpose of investigating the effects of ${{\mathbf{E}}^0}$ and ${{\mathbf{B}}^c}$ on the $\textrm{ICD}$, we change the structure  to regulate these factors. One additional hole is set at the center of every four adjacent original holes.

The radius of additional holes is changed from 20 $nm$ to 100 $nm$ by step of 20 $nm$. \hyperlink{ex:ebeff}{Figure} \ref{ebeff} shows the charge distribution at three modes with changing of ${R_\alpha }$. From \hyperlink{ex:ebeff} {figure} \ref{ebeff} it can be seen that SPR strength gradually decreases at mode \uppercase\expandafter{\romannumeral1}, and increases at mode \uppercase\expandafter{\romannumeral2} and \uppercase\expandafter{\romannumeral3}. Thereby, the electric field strength ${{\mathbf{E}}^{0}}$ and magnetic flux density ${{\mathbf{B}}^{c}}$ decreases at mode \uppercase\expandafter{\romannumeral1}, and increases at mode \uppercase\expandafter{\romannumeral2} and \uppercase\expandafter{\romannumeral1}\,.
\renewcommand\thefigure{S\arabic{figure}}
\begin{figure}[h]
\centering
\includegraphics[height=7cm]{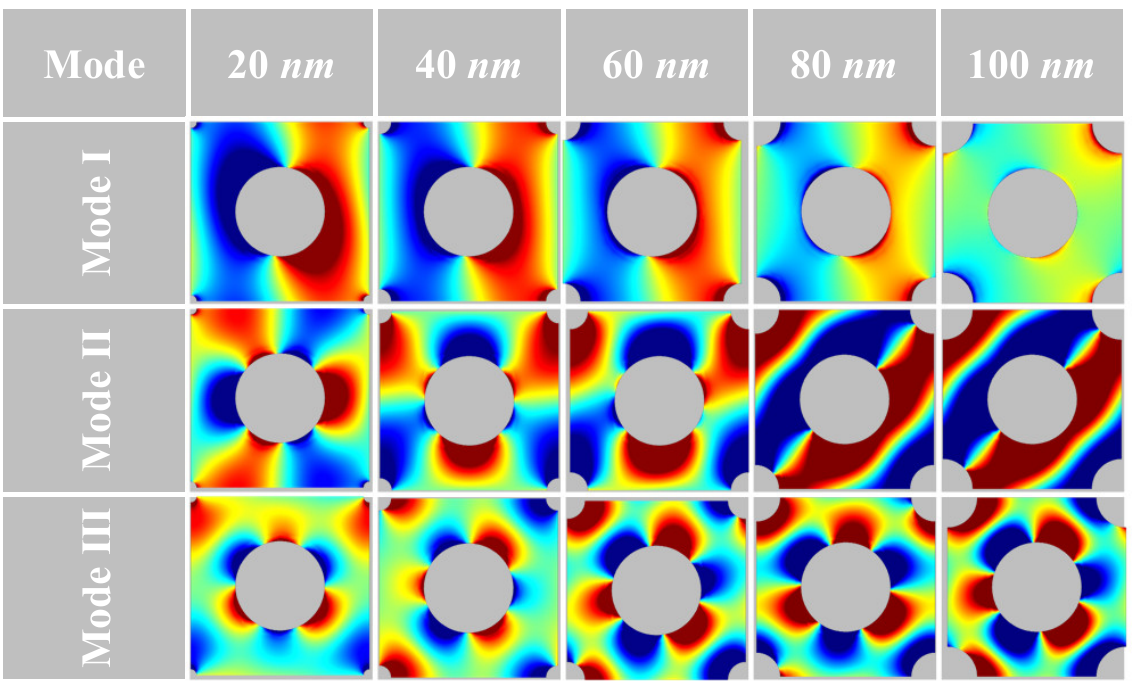}
\hypertarget{ex:ebeff}{\caption{The charge distribution of GNM with different radius ${R_\alpha }$.}}
\label{ebeff}
\end{figure}

\section{ Calculation of chiral medium by FEM}
 We used COMSOL software for the simulation. For calculation of chiral medium we changed part of equations through implementing chiral constitutive relation \cite{SCKelly}\,.  To confirm our calculation results, we applied these modifications to solve for the transmission coefficients for a chiral slab and compared the simulation results of the co-and cross-components of the transmission coefficients at normal incidence from a slab with thickness $L$. The results are:
\renewcommand\theequation{S\arabic{equation}}
\begin{eqnarray}
 \textrm{T}_{co} &=& \frac{2\eta \eta_0 cos(kL)cos(\kappa kL)}{2\eta \eta_0 cos(kL)+i(\eta_0^2+\eta^2)sin(kL)}\\
 \textrm{T}_{cr} &=& \frac{2\eta \eta_0 cos(kL)sin(\kappa kL)}{2\eta \eta_0 cos(kL)+i(\eta_0^2+\eta^2)sin(kL)}\nonumber
\end{eqnarray}

 Figure \ref{figeslab} depicts infinite chiral slab in air and calculation result. The simulated and analytical results are in a good agreement.
\renewcommand\thefigure{S\arabic{figure}}
\begin{figure}[h]
\centering
\includegraphics[height=4.5cm]{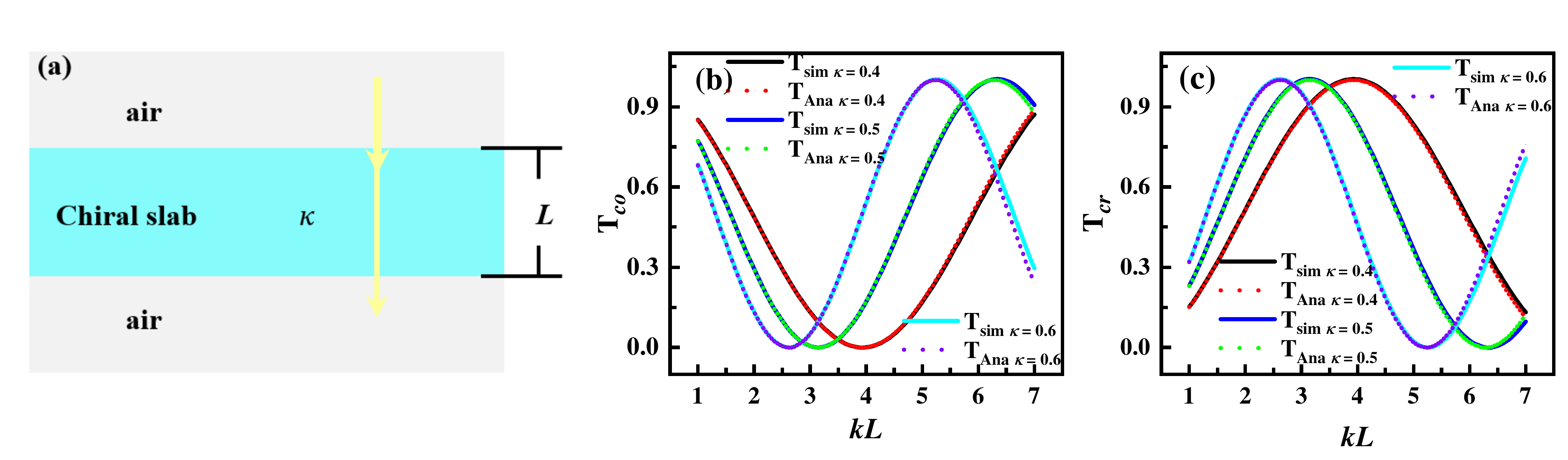}
\caption{(a) Chiral slab (b) Analytical (dashed line) and simulated (solid line) co-component of transmission (c) Analytical (dashed line) and simulated (solid line) cross-component of transmission}
\label{figeslab}
\end{figure}

For anisotropic chiral medium, the constitutive relations are expressed by tensorial chirality parameter. The unitary transformations are used for rotation of orientation, for example, for rotation about $y$ axis the constitutive relations are modulated as
\renewcommand\theequation{S\arabic{equation}}
\small
\begin{eqnarray}
\left(\begin{array}{ccc}
\textrm{D}^x\\
\textrm{D}^y\\
\textrm{D}^z\\
\end{array}\right)&=&\varepsilon_{0}\varepsilon_{c}\left(\begin{array}{ccc}
\textrm{E}^x\\
\textrm{E}^y\\
\textrm{E}^z\\
\end{array}\right)\nonumber\\
&-&\frac{i}{c}
\left(\begin{array}{ccc}
cos(\alpha)&0&-sin(\alpha)\\
0&1&0\\
sin(\alpha)&0&cos(\alpha)\\
\end{array}\right)
\left(\begin{array}{ccc}
\kappa_{xx}&\kappa_{xy}&0\\
\kappa_{yx}&\kappa_{yy}&0\\
0&0&\kappa_{zz}\\
\end{array}\right)
\left(\begin{array}{ccc}
cos(\alpha)&0& sin(\alpha)\\
0&1&0\\
-sin(\alpha)&0&cos(\alpha)\\
\end{array}\right)
\left(\begin{array}{ccc}
\textrm{B}^x\\
\textrm{B}^y\\
\textrm{B}^z\\
\end{array}\right)
\end{eqnarray}

\begin{eqnarray}
\left(\begin{array}{ccc}
\textrm{B}^x\\
\textrm{B}^y\\
\textrm{B}^z\\
\end{array} \right)&=&\mu_{0}\mu_{c}\left(\begin{array}{ccc}
\textrm{H}^x\\
\textrm{H}^y\\
\textrm{H}^z\\
\end{array}\right)\nonumber\\
&+&\frac{i}{c}
\left(\begin{array}{ccc}
cos(\alpha)&0&-sin(\alpha)\\
0&1&0\\
sin(\alpha)&0&cos(\alpha)\\
\end{array}\right)
\left(\begin{array}{ccc}
\kappa_{xx}&\kappa_{xy}&0\\
\kappa_{yx}&\kappa_{yy}&0\\
0&0&\kappa_{zz}\\
\end{array}\right)
\left(\begin{array}{ccc}
cos(\alpha)&0& sin(\alpha)\\
0&1&0\\
-sin(\alpha)&0&cos(\alpha)\\
\end{array}\right)
\left(\begin{array}{ccc}
\textrm{E}^x\\
\textrm{E}^y\\
\textrm{E}^z\\
\end{array} \right)
\end{eqnarray}
\normalsize
\\

\end{document}